\documentclass[journal]{IEEEtran}
\usepackage[T1]{fontenc}
\usepackage{float}
\usepackage{amsmath}
\usepackage{amsthm}
\usepackage{amssymb}
\usepackage{stackrel}
\usepackage{graphicx}
\usepackage{setspace}
\usepackage{url}
\allowdisplaybreaks

\usepackage{caption}
\usepackage{subfigure}
\usepackage{algorithm}
\usepackage[graphicx]{realboxes}
\usepackage{algorithmic}
\usepackage{verbatim}
\renewcommand{\algorithmicensure}{ \textbf{Output:}} 
\renewcommand{\algorithmicrequire}{ \textbf{Initialize:}} 
\makeatletter

\floatstyle{ruled}
\newfloat{algorithm}{tbp}{loa}
\providecommand{\algorithmname}{Algorithm}
\floatname{algorithm}{\protect\algorithmname}
\theoremstyle{plain}

\theoremstyle{definition}

\theoremstyle{plain}

\theoremstyle{plain}
\newcommand{\RNum}[1]{\uppercase\expandafter{\romannumeral #1\relax}}
\usepackage[thicklines]{cancel}
\IEEEoverridecommandlockouts
\usepackage{ifpdf}
\usepackage{cite}
\usepackage{graphicx}
\ifCLASSOPTIONcompsoc
\else
\fi
\usepackage{graphicx}
\usepackage{epstopdf}
\usepackage{amsmath}
\usepackage{mathtools}
\usepackage{booktabs} 
\usepackage{multirow}
\usepackage{setspace}
\usepackage{lettrine} 
\usepackage{bm}
\makeatother
\usepackage{babel}
\newtheorem{define}{Definition}

\newtheorem{lemm}{Lemma}

\newtheorem{prop}{Proposition}

\usepackage{color}
\usepackage{booktabs} 
\usepackage{longtable}
\usepackage{amsthm} 

\begin{document}
\captionsetup[figure]{font={small}, name={Fig.}, labelsep=period}
\title{Multi-Objective Optimization-Based Waveform Design for  Multi-User and Multi-Target MIMO-ISAC Systems}
	\author{
		Peng Wang,
		Dongsheng Han,~\IEEEmembership{Member,~IEEE},
		Yashuai Cao,
		Wanli Ni,~\IEEEmembership{Member,~IEEE},\\
		and 
		Dusit Niyato,~\IEEEmembership{Fellow,~IEEE}
\thanks{	
		This research is supported in part by the S$\&$T Program of Hebei under Grant SZX2020034, and in part by the Project funded by China Postdoctoral Science Foundation under Grant No. 2023M742009, and in part by the National Research Foundation, Singapore, and Infocomm Media Development Authority under its Future Communications Research $\&$ Development Programme, Defence Science Organisation (DSO) National Laboratories under the AI Singapore Programme (AISG Award No: AISG2-RP-2020-019 and FCP-ASTAR-TG-2022-003), and Singapore Ministry of Education (MOE) Tier 1 (RG87/22).
	}	
\thanks{
	The conference version of this paper was accepted in ICASSP 2024~\cite{conf1}.
}	
	\thanks{
	P. Wang and D. Han are with the School of Electrical and Electronic Engineering, North China Electric Power University, Beijing 102206, China, and also with Hebei Key Laboratory of Power Internet of Things Technology, North China Electric Power University, Baoding 071003 China (e-mail: \{wangpeng9712, handongsheng\}@ncepu.edu.cn).
}
	\thanks{
	Y. Cao and W. Ni are with the Department of Electronic Engineering, Tsinghua University, Beijing 100084, China (e-mail: \{caoys, niwanli\}@tsinghua.edu.cn).
}
\thanks{
    D. Niyato is with the School of Computer Science and Engineering, Nanyang Technological University, Singapore 639798 (e-mail: {dniyato}@ntu.edu.sg).
}
}
\maketitle
\begin{abstract}
	Integrated sensing and communication (ISAC) opens up new service possibilities for sixth-generation (6G) systems, where both communication and sensing (C$\&$S) functionalities co-exist by sharing the same hardware platform and radio resource. In this paper, we investigate the waveform design problem in a downlink multi-user and multi-target ISAC system under different C$\&$S performance preferences. The multi-user interference (MUI) may critically degrade the communication performance. {To eliminate the MUI, we employ the constructive interference mechanism into the ISAC system, which saves the power budget for communication. However, due to the conflict between C$\&$S metrics, it is intractable for the ISAC system to achieve the optimal performance of C$\&$S objective simultaneously. Therefore, it is important to strike a trade-off between C$\&$S objectives. }By virtue of the multi-objective optimization theory, we propose a weighted Tchebycheff-based transformation method to re-frame the C$\&$S trade-off problem as a Pareto-optimal problem, thus effectively tackling the constraints in ISAC systems. Finally, simulation results reveal the trade-off relation between C$\&$S performances, which provides insights for the flexible waveform design under different C$\&$S performance preferences in MIMO-ISAC systems.  	 
\end{abstract}	

\section{Introduction}
 \IEEEPARstart{D}{ue} to the dual capability of communication and radar sensing, integrated sensing and communication (ISAC) has inspired enormous academic and industrial enthusiasm \cite{feng2021joint,zhang2021overview}. The appealing concept stems from the fact that communication and radar systems share many similarities in terms of devices, waveform, and signal processing algorithms\cite{chen2021code}. What's more, the overlapping of radar and communication frequency bands developing towards higher frequency such as millimeter wave, terahertz and visible light, promotes the interoperability of communication and sensing (C$\&$S) with existing wireless infrastructure\cite{mishra2019toward}. Consequently, ISAC has been identified as a promising technique to alleviate the exacerbated spectrum scarcity for the upcoming sixth-generation (6G) wireless systems. Triggered by benefits from ISAC, a unified theoretical framework for ISAC resource allocation is likely to drive a paradigm innovation for future cellular architecture and networking protocols.\par
 Compared to the dedicated sensing or communication framework, ISAC not only utilizes hardware, software, and wireless resources efficiently, but also achieves mutual enhancement between dual functions of C$\&$S \cite{zhang2021enabling}. Integrated waveform is fundamental and crucial to achieving dual functions of C$\&$S, which brings two gains to ISAC system: integration gain and coordination gain\cite{cui2021integrating}. However, the integrated waveform design for ISAC systems is currently confronted with critical challenges\cite{liu2022survey,barneto2019full}. For example, 1) the integrated waveform needs to be designed flexibly and adjusted dynamically according to the actual requirements of ISAC scenarios and environmental changes. 2) The coexistence of C$\&$S functions at the same time and frequency creates conflict and interference. As such, it is highly desired to develop a flexible integrated waveform design strategy adapting to the demanding requirements of actual ISAC system under severe interference.\par

\subsection{Related Work}
Some prior studies have investigated the integrated waveform design in ISAC systems, where the waveform is designed according to three principles: sensing-centric\cite{5393298}, communication-centric\cite{9399801}, and joint design\cite{cui2021integrating}. For the sensing-centric scheme, ISAC empowered radars with communication capabilities by embedding communication data into existing sensing waveforms\cite{huang2020majorcom}. To achieve the ``dual-use'' of radar waveforms, a classic radar-centric ISAC waveform design involved utilizing distinct sensing waveforms during coherent processing intervals to represent various broadcast communication symbols in \cite{blunt2010embedding}. Based on this approach, the authors of \cite{blunt2010intrapulse} designed a radar pulse compression filtering framework to mitigate the degradation of radar anti-interference performance caused by transmit diversity. Nevertheless, sensing-centric design was only applicable to limited communication scenarios due to its lower transmission rate. On the contrary, communication-centric scheme refered to the direct utilization of existing communication waveform for radar sensing, in which the communication signal acted as a radar probing signal for target location, detection, and tracking target\cite{kumari2017ieee}. For this scheme, communication waveforms need to be carefully shaped to meet specific sensing constraints, such as low peak-to-average power ratio (PAPR), desirable correlation properties and robustness against interference\cite{barneto2019ofdm,barneto2019full}. However, due to the fact that communication waveforms were not tailored specifically for radar applications, their sensing ability was difficult-to-tune, requiring sophisticated signal processing to compensate for performance degradation.\par
{Both radar-centric and communication-centric schemes were inflexible and severely scenario-dependent. To extend ISAC to more practical applications, an integrated waveform based on the joint design of radar and communication was proposed, where sensing and communication-centric schemes can be viewed as two extreme cases \cite{dong2020low}. The joint design for integrated waveform can be formulated as a mathematical optimization problem to simultaneously consider the performance metrics for C$\&$S functionalities \cite{tian2021transmit,shi2020joint,9124713,10086626}. The authors of \cite{9124713} focused on the weighted sum of the beam pattern and the cross correlation pattern by jointly precoding individual communication and radar signals. To eliminate interference caused by radar signals before decoding the information signal, an improved interference cancellation receiver was designed in \cite{10086626}.}\par 
{However, C$\&$S metrics in the ISAC system were not aligned, making it impossible to achieve the optimal performance of C$\&$S. Therefore, it is necessary to traded off performance of C$\&$S for designing an integrated waveform. In recent years, a large number of research have emerged, which tarde off the performance of ISAC systems and provide performance boundaries based on different metrics \cite{10032141}, \cite{9618653}. To be specific, the authors in \cite{10217169} investigated the fundamental performance trade-off in a point-to-point MIMO-ISAC system where both point and extended target models were considered, The Pareto boundary in terms of communication rate and sensing CRB was analyzed. The authors of \cite{xiong2023fundamental} and \cite{liu2021cramer} analyzed the performance limits and achievable performance boundaries of ISAC based on information theory, while investigating the trade-off between C$\&$S. Nevertheless, these works rarely consider the performance trade-off between C$\&$S by quantifying the effects of different system setting factors on the system performance. In fact, system setting factors, such as the number of users, targets and transmit antennas, potentially affect the Pareto boundary \cite{9468975}. Especially, the trade-off is significantly influenced by the transmit power due to the fact that equal power allocation results in unequal performance gains of sensing/communication \cite{9965407}, \cite{9945983}. Consequently, investigating the influence of diverse factors on the performance trade-off is crucial for designing ISAC waveforms.}\par

 \subsection{Motivations and Challenges}
The motivations of this paper include three aspects. \emph{Firstly}, the single-user or single-target ISAC network has been comprehensively studied in \cite{ren2023robust,su2020secure,su2022secure,hua2023secure} because of simplified models and analysis. Compared to the single-user and single-target ISAC system, multi-user ISAC system is more realistic. On the other hand, multi-user ISAC system is rather challenging because the coupling between C$\&$S becomes more complicated to deal with in the presence of multi-user interference (MUI). Intuitively, MUI can lead to a decrease in the communication rate for users. To meet the communication requirements of ISAC system under a given power budget, partial transmit power originally used for sensing is allocated to users for communication, which indirectly degrades the sensing performance.\par
 \vspace{1mm}
\emph{Secondly}, the performance requirements of C$\&$S are different in diverse application scenarios, and even change within the same scenario. However, traditional ISAC waveform design approaches are difficult to adapt to the various design objectives. It is urgent to seek a general optimization framework to address the challenge of allocating and utilizing wireless resources efficiently and flexibly, considering the varying performance preferences.\par
\emph{Thirdly}, to achieve the optimal performance for the ISAC system, it is necessary to trade-off C$\&$S performances since two involved objectives are conflicting. This trade-off becomes particularly vital in multi-user and multi-target scenarios, because the performance constraint of each user and target needs to be considered. This further increases the degree of coupling between C$\&$S performances.  After all, fulfilling these constraints requires careful coordination of the individual devices just like a game-playing. \par
 \begin{figure*}[t]
 	\centering{}\includegraphics[width=7.1in]{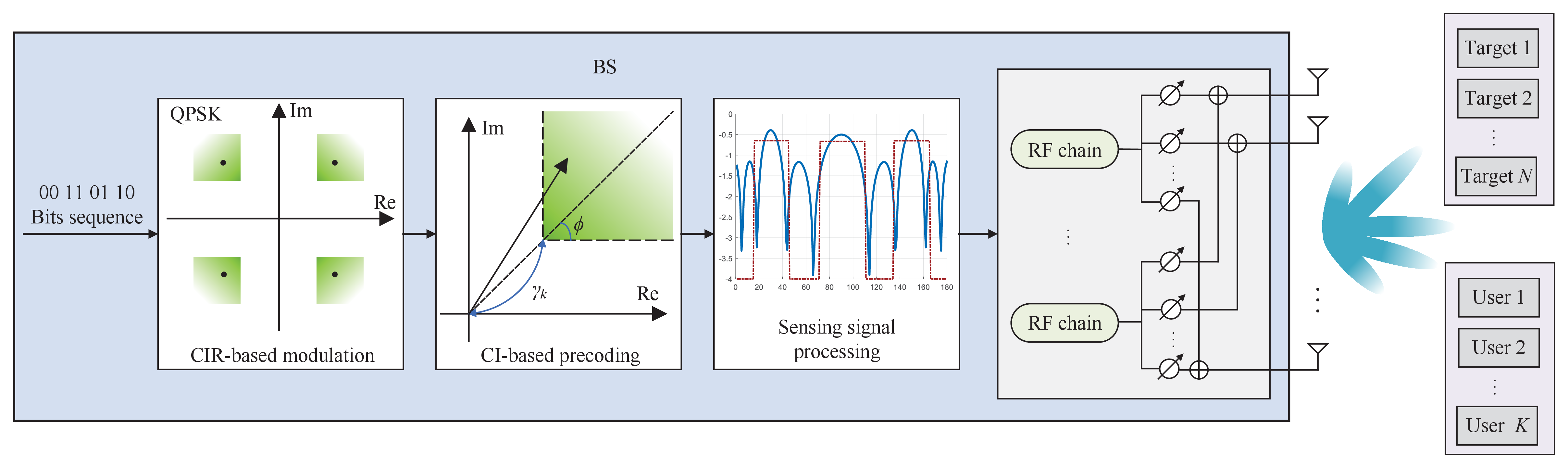}
 	\caption{CI-based multi-user and multi-target ISAC framework.}
 	\label{fig:signal_processing}
 \end{figure*}

 \subsection{Contributions and Organization}
 {To overcome the aforementioned challenges, we devise a general integrated waveform optimization framework for multi-user and multi-target ISAC systems with different C$\&$S performances. To this end, we formulate the joint waveform design problem as a MOOP, and propose a weighted Tchebycheff-based transformation method to trade off C$\&$S performances. Based on the obtained Pareto front, we investigate the impact of transmitter parameters on the C$\&$S performances. The key contributions of the paper are summarized as follows.}\par
 \begin{itemize}
 	\item[$\bullet$] {To eliminate the MUI in the multi-user and multi-target ISAC systems, we employ the constructive interference (CI) principle to the ISAC system. With the developed scheme, known MUI can be transformed into useful signals by CI-based precoding, which can also reduce the power budget for communication. Thus, more power can be utilized to improve the sensing ability.}
 \end{itemize}
 \begin{itemize}
 	\item[$\bullet$] {We formulate the joint waveform design problem as a MOOP whose objective function is a vector form with conflicting relationship. To trade off the C$\&$S performances, we propose a weighted Tchebycheff-based transformation method to reformulate the MOOP into a Pareto-optimization problem.} 
 \end{itemize}
 \begin{itemize}
 	\item[$\bullet$] {We propose an efficient successive convex approximation (SCA) algorithm to tackle the formulated non-convex MOOP, where geometric programming (GP) and semi-definite relaxation (SDR) methods are adopted to solve the decoupled sub-problems. We also present numerical results to reveal the impact of transmitter parameters on the C$\&$S performances, which provides valuable insights for facilitating the integrated  waveform design.}
 \end{itemize}

\section{System Model}\label{2}
In this paper, we consider a multiple-input multiple-output (MIMO) ISAC system consisting of one dual-functional radar-communication (DFRC) base station (BS) equipped with a half-wavelength uniform linear array (ULA) with $N_t$ transmit antennas and $N_r$ receive antennas, $K$ single-antenna communication users and $N$ point-like sensed targets. The set of communication users and the set of sensed targets are indexed by $\mathcal{K}=\left\{1,2,...,K\right\}$  and $\mathcal{N}=\{1,2,...,N\}$, respectively. All communication users served  by the downlink broadcast channels, and the DFRC BS is used to communicate with $K$ single-antenna users and sense $N$ point-like targets, simultaneously\footnote{{In ISAC scenarios, both periodic and continuous sensing designs are prevalent \cite{9858656}, \cite{9124713}. In this paper, the continuous sensing design is considered for delay-sensitive services, such as autonomous driving scenarios.}}.  
We use the vector $\mathbf{s}=\left[s_1,s_2,...,s_K\right]^T\in\mathbb{C}^{K\times1}$ to collect the transmit symbols, where $s_k$ denotes the intended symbol $M$-PSK modulated of the $k$-th user and $s_k\in\{e^{\frac{j\pi}{M}},e^{\frac{j3\pi}{M}},...,e^{\frac{j(2M-1)\pi}{M}}\}$. The transmit signal at the BS can be expressed as
\begin{align}
\mathbf{x}=\mathbf{Ws}=\sum_{k=1}^K\mathbf{w}_ks_k,
\end{align}%
where $\mathbf{W}=[\mathbf{w}_1,\mathbf{w}_2,...,\mathbf{w}_K]\in\mathbb{C}^{N_t{\times}K}$ is the precoding matrix and $\{\mathbf{w}_k\}_{k=1}^K \in \mathbb{C}^{N_t\times1}$ is the beamforming vector of the $k$-th users.\par

\subsection{Communication Model}
\vspace{2mm}
For the communication, the received signal at the $k$-th user can be written as
\begin{align}
y_k=\mathbf{h}_k^H\mathbf{x}+n_k, \forall{k} \in \mathcal{K}, \label{y_k}
\end{align}\vspace{0.2em}%
where $\mathbf{h}_k\in\mathbb{C}^{N_t{\times}1}$ denotes the multiple-input single-output (MISO) channel vector between the BS and the $k$-th user, and $n_k\sim\mathcal{CN}(0,N_0),\forall{k} \in \mathcal{K}$ is the noise at the receiver of the $k$-th user, that obeys circularly symmetric complex Gaussian distribution with mean zero and variance $N_0$. \par
We assume that there are line-of-sight (LoS) paths and a large number of scattering points between the BS and users in the considered scenario. Therefore, we consider that the channel $\mathbf{h}_k$ is subject to slow time-varying block Rician fading. This means that the channel remains constant within a block but varies slowly from one block to the next. Then, we can express the channel vector of the $k$-th user as a composite of two components: a line-of-sight (LoS) channel vector, representing the strongest signal path, and a non-line-of-sight (NLoS) channel vector resulting from multiple-path scattering, which is given by
\begin{align}
\mathbf{h}_k=\sqrt{\frac{v_k}{1+v_k}}\mathbf{h}_k^{\mathrm{LoS}}+\sqrt{\frac{1}{1+v_k}}\mathbf{h}_k^{\mathrm{NLoS}}, \forall{k} \in \mathcal{K},
\end{align}\vspace{0.3em}%
where $v_k>0$ is the Rician factor of the $k$-th user, $\mathbf{h}_k^{\mathrm{LoS}}$ is the LoS component and $\mathbf{h}_k^{\mathrm{NLoS}}$ is the NLoS component. \par
From \eqref{y_k}, the signal-to-interference-plus-noise ratio (SINR) at the $k$-th user is 
\begin{align}
\gamma_k=\frac{\left|\mathbf{h}_k^H\mathbf{w}_k\right|^2}{\sum_{i=1,i{\ne}k}^K\left|\mathbf{h}_k^H\mathbf{w}_i\right|^2+N_0}, \forall{k} \in \mathcal{K}. \label{sinr}
\end{align}
In our proposed system model, DFRC BS serves both target sensing and user communication simultaneously, creating coupling in power. Given a fixed transmit power budget for the system, we introduce CI mechanism into the model to eliminate the MUI in multi-user and multi-target scenarios as Fig. \ref{fig:signal_processing}. Different from traditional precoding techniques that aim to reduce bit error rate (BER) by confining the received signal within the vicinity of the constellation point for users, CI-based precoding restricts the received signal of each user to a specific region known as the constructive interference region (CIR). By following the principle of CI\cite{masouros2015exploiting}, MUI will not be suppressed, but will be regarded as a beneficial signal source, which also reduces the need for communication to transmit power. To be specific, by the availability of channel information and intended communication data from all users, the phases of interference can be aligned with the desired symbol through implementation of beamforming vector $\{\mathbf{w}_i\}_{i=1}^K$. This alignment enables constructive superposition of the signals, allowing the effective utilization of interference as a useful signal source\footnote{{In the proposed ISAC system, robust time synchronization can be established by the existing deliberate synchronized constructive interference method \cite{6774898}.}}. \par
 Accordingly, for the $M$-PSK modulation based on CI\footnote{{We consider the M-PSK modulation for communication due to its characteristic of constant amplitude. In fact, CI can also be extended to QAM modulation  by introducing additional amplitude alignment constraints.}}, the SINR at the $k$-th user in \eqref{sinr} can be rewrite as  
\begin{align}
	\gamma_k=\frac{\left|\mathbf{h}_k^H\sum_{i=1}^K\mathbf{w}_is_i\right|^2}{N_0},\forall{k} \in \mathcal{K}.
\end{align}
Based on the Shannon capacity theory, the downlink communication rate of the $k$-th user is given by
\begin{align}
	R_k=B\log_2\left(1+\frac{\left|\mathbf{h}_k^H\mathbf{x}\right|^2}{N_0}\right),\forall{k} \in \mathcal{K},
\end{align}
where $B$ is the available bandwidth at the BS. Hence, the downlink sum rate of all  communication users in ISAC system is $R_{\mathrm{sum}}=\sum^K_{k=1} R_k$.
\vspace{-4mm}
\subsection{Radar Sensing Model}
For the radar sensing, it is crucial to employ an energy-focused beam with low sidelobe leakage to illuminate the desired sensing location\cite{jardin2010wideband}. This ensures that the desired echoes can be easily distinguished from clutter, allowing for effective separation between the desired signals and undesired interference. 
The beampattern can clearly convey the gain characteristics of the beam in all directions.
To this end, in order to be able to flexibly shape the transmitted signal to achieve the desired beam characteristics, we design beamforming to obtain the desired beampattern by manipulating the covariance of the transmit signals.  
The beampattern of the transmit signal $\mathbf{x}\in\mathbb{C}^{N_t{\times}1}$ is expressed as
\begin{align}
	G(\theta)=\mathbb{E}\left\{\left|\mathbf{a}_t^H(\theta)\mathbf{x}\right|^2\right\}=\mathbf{a}_t^H(\theta)\mathbf{R}\mathbf{a}_t(\theta),\label{7}
\end{align}
where $\mathbf{a}_t(\theta)$ is the steering vector of the transmit antenna array, and $\mathbf{R}=\mathbf{x}\mathbf{x}^H$ is the covariance matrix of the transmit signal. Then, the steering vector of the transmit and receive antennas can be shown below, respectively.
\begin{align}
	\mathbf{a}_t(\theta)=\frac{1}{\sqrt{N_t}}\left[1,e^{-j\pi\sin\theta},...,e^{-j\pi(N_t-1)\sin\theta}\right]^T, \\
	\mathbf{a}_r(\theta)=\frac{1}{\sqrt{N_r}}\left[1,e^{-j\pi\sin\theta},...,e^{-j\pi(N_r-1)\sin\theta}\right]^T.
\end{align}
 
Therefore, the received echoes from targets at receive antennas are denoted by
 {
\begin{align}
\mathbf{y}_r= \sum\nolimits^N_{n=1}\beta_n\mathbf{a}_r(\theta_n)\mathbf{a}_t^T(\theta_n)\mathbf{x}+\mathbf{z},
\end{align}}%
where $\beta_n\in\mathbb{C}$ denotes the complex  {round-trip} channel coefficient of $n$-th target depending on the associated path loss and its radar cross section, $\mathbf{z}\in\mathbb{C}^{N_r{\times}1}$ denotes the noise at the BS receive antenna (including clutters and interference), and $\theta_n$ is the angle of the $n$-th target. 
\vspace{-3mm}
\section{Problem Formulation}\label{3}
In this section, based on the multi-objective optimization theory, we propose a flexible integrated waveform design scheme for multi-user and multi-target ISAC systems. To this end, we firstly formulate two single-objective optimization problems for maximizing the communication sum rate in constraints of CIR and minimizing the radar sensing error with the knowledge of precise target location. Then, we formulate the C$\&$S performances trade-off problem as the MOOP for achieving optimal ``dual-function'' under scenarios with different C$\&$S performance preferences.
\vspace{-5mm} 
\subsection{Precoding Design with CI}\label{2.a}
\vspace{-1mm}
 {In conventional optimizations, the sum rate is generally optimized subject to SINR constraints as problem \eqref{csr}. 
\begin{subequations}
\begin{align}
 		\underset{\{\mathbf{w}_k\}}{\max}\quad&\sum_{k=1}^KB\log_2\left( 1+\gamma_k \right) \\
 		s.t.\quad&\frac{|\mathbf{h}_k^T\mathbf{w}_k|^2}{\sum_{i=1,i\ne k}|\mathbf{h}_k^T\mathbf{w}_i|^2+N_0}\ge \Gamma_k, \forall k, \label{11b}\\
 		&||\mathbf{x}||^2\le P_{\max}.
\end{align}\label{csr}%
\end{subequations}%
where all interference is treated as harmful and constrained by the given threshold $\Gamma_k$, so that the receiver can decode symbols that are within a certain distance from the nominal constellation. In fact, some interference such as MUI can be treated as beneficial by optimizing the precoding matrix $\mathbf{w}_k$\cite{masouros2015exploiting}. This approach also allows us to lower the power consumption requirements for communication, thereby freeing up more power for sensing. }\par
{Following the advantage of CI technique against MUI, we artificially construct a CIR by imposing constraints during modulating and precoding. The details of CIR-based modulation and CI-based precoding modules are shown in Fig. \ref{fig:signal_processing}. In CIR, the MUI will constructively interfere with each other, resulting in a gain that is beneficial to the signal. Given the availability of channel information and intended communication data from all users, we can re-establish the constraint \eqref{11b} as 
 \begin{align}
 \left|\mathrm{angle}\{y_k-n_k\}-\mathrm{angle}\{s_k\}\right| \le \Phi, \forall{k} \in \mathcal{K},\label{CI1}
\end{align}
  \begin{align}
\frac{|\mathbf{h}_k^T\sum_{i=1}^K\mathbf{w}_is_i|^2}{N_0} \ge \Gamma_k, \forall{k} \in \mathcal{K},\label{CI2} 
\end{align}
 where $\Phi$ is tolerable phase difference between the noise-free received symbols and the intended symbol of the $k$-th communication user, $\Gamma_k$ is the given SNR threshold, and $P_{\max}$ is the maximal transmit power.}\par 
 {As illustrated in Fig. \ref{fig:signal_processing}, taking QPSK modulation as an example, the green region represents the CIR. To facilitate later optimizations, we rotate the constructed CIR by $\mathrm{angle}\{s_k^*\}$, then constraints \eqref{CI1} and \eqref{CI2} can be equivalent to
 \begin{align}
 	\left|\mathrm{Im}(\hat{y}_ks_k^*)\right|\le\left(\mathrm{Re}\left(\hat{y}_ks_k^*\right)-\sqrt{N_0\Gamma_k}\right)\tan\phi,\forall{k} \in \mathcal{K}, 
 \end{align}}%
 where $\hat{y}_k =y_k - n_k$ is the noise-free received symbols, $\phi=\pm \pi/M$ denotes the maximum angle shift in the CIR. Thus, for the $M$-PSK modulation based on CI, the problem of maximizing achievable sum rate for communication under the communication users' quality of service (QoS) and transmit power budget constraints is reformulated as
\begin{subequations}
\begin{align}
\mathrm{SOOP 1}:\quad\underset{{\mathbf{x}}}{\max}\quad
     &R_{\mathrm{sum}}  \label{SOOP1a}\\ 
\mathrm{s.t.}\quad
	&\left|\mathrm{Im}\left(\mathbf{h}_k^H\mathbf{x}s_k^*\right)\right|\nonumber\\ 
	\le\mathrm{Re}\left(\mathbf{h}_k^H\mathbf{x}s_k^*\right)&\tan\phi-\sqrt{N_0\Gamma_k}\tan\phi,\forall{k} \in \mathcal{K} \label{SOOPb}, \\ 
	&||\mathbf{x}||^2 \le P_{\max} \label{13c}, 
\end{align}
\label{SOOP1}%
\end{subequations}
where constraint \eqref{SOOPb} indicates that the signal constellation point received by the $k$-th user must fall in the CIR geometrically, and the maximum signal power is less than the transmit power budget in constraint (\ref{13c}).

\subsection{MIMO Radar Waveform Design}\label{2.b}
\vspace{2mm}
In the radar sensing process, MIMO radar first obtains the rough angle information of the target through the wide beam in the detection stage, and then sends the sensing signal to track the target. In this paper, our focus is directed towards the tracking phase, where the objective is to enhance the probing gain specifically towards the angles of interest. Towards this end, we adopt the mean squared error (MSE) as our performance metric of radar sensing error to assess the quality of beampattern matching. The MSE quantifies the difference between the actual transmit beampattern $G(\theta)$  and the desired beampattern $\hat{G}(\theta)$ in the angular domain, providing a measure of the effectiveness of the beampattern matching. {The beampattern matching MSE is expressed as}\par
	{\begin{align}
		&M_s\left(\eta,\mathbf{R}\right)=\frac{1}{L}\sum^L_{l=1}\left|\eta\hat{G}(\theta_l)-\mathbf{a}^H(\theta_l) \mathbf{R} \mathbf{a}(\theta_l)\right|^2, \label{14}
	\end{align}}%
{where $\theta_l$ denotes the $l$-th sample angles over $[-\frac{\pi}{2},\frac{\pi}{2}]$, and $\eta$ is a scaling coefficient.}\par
\vspace{1mm}
{Note that it is essential to control the transmit beampattern $G(\theta)$ to mimic the desired transmit beampattern $\hat{G}(\theta)$ based on the estimated direction $\hat{\theta}_n$ \cite{kumari2021adaptive}. Therefore, a desired beampattern $\hat{G}(\theta)$ must be defined prior to designing the transmit waveform, which is given by}\par
{
	\begin{align}
		\hat{G}(\theta)=
		\begin{cases}
			1,\quad|\theta-\hat{\theta_n}|<\frac{\Delta\theta}{2},n=1,2,...,N, \\
			0,\quad \mathrm{otherwise},\\
		\end{cases}
	\end{align}
	where $\Delta\theta$ denotes the width of desired beampattern at each estimation angle.}\par  
	\vspace{1mm}
Herein, our objective is to optimize the covariance matrix $\mathbf{R}$ of the ISAC waveform to minimize the beampattern matching MSE between the actual transmit beampattern and the desired beampattern in \eqref{14}, while ensuring compliance with the maximal power budget at the DFRC BS. The problem of minimizing sensing error can be formulated as
	\begin{subequations}
		\begin{align}
			\mathrm{SOOP 2}:\quad\underset{\eta,\mathbf{R}}{\min}\quad
			&M_s\left(\eta,\mathbf{R}\right)\label{16a}\\ 
			\mathrm{s.t.}\quad	
			&\mathrm{Tr}(\mathbf{R}) \le P_{\max},\label{16b}\\
			&\mathbf{R}\succeq 0 \label{16c},\\
			&\mathrm{rank}\{\mathbf{R}\}=1.\label{16d} 
		\end{align}
		\label{SOOP2}%
\end{subequations}
{In the SOOP2, we obtain the desired beampattern by optimizing the variable $\mathbf{R}$, which is essentially an optimization of the waveform $\mathbf{x}$ since $\mathbf{R}$ is the covariance matrix of $\mathbf{x}$ that satisfies $\mathbf{R}=\mathbf{x}\mathbf{x}^H$. Therefore, constraints \eqref{16c} and \eqref{16d} are added to SOOP2 to ensure that the variable $\mathbf{R}$ is a semidefinite matrix.}

\subsection{Multi-Objective Optimization for ISAC}

In the above two subsections, SOOP1 and SOOP2 are formulated to realize communication sum rate maximization and sensing error minimization, respectively. It is obvious that the two optimization problems do not coincide with each other. The variables $\mathbf{R}$ and $\mathbf{x}$ are coupled based on the fact that $\mathbf{R}$ is the covariance matrix of $\mathbf{x}$. From the perspective of optimization, this coupling directly leads to a conflicting relationship between SOOP1 and SOOP2 in terms of transmit power, as governed by constraints \eqref{13c} and \eqref{16b}. Therefore, it is a challenge to optimize both objectives from SOOP1 and SOOP2, simultaneously. In addition, the impact of this coupling on C$\&$S performances may also be reflected in other aspects, such as the number of transmit antennas, which will be discussed in Section \ref{5}. In this subsection, we aim to simultaneously optimize the performances of C$\&$S in the proposed ISAC system.   \par
\vspace{1mm}
In conventional waveform designs for ISAC system, single-objective optimization-based schemes focus on one single performance (e.g. communication or sensing) metric, while the other performance metric only acts as a threshold constraint. Such design  only attempt to achieve the optimal performance of one single metric with limited resources, and just meet the basic threshold requirements for other performance metrics. Additionally,  traditional single-objective optimization only offers an optimal solution under the given constraints, thus reducing the flexibility of the integrated system design. Therefore, a more fair and practical optimization should satisfy multiple objectives simultaneously, such as power budget, sensing accuracy, and communication rate, or trade-off among the above objectives. Multi-objective optimization-based scheme can jointly consider C$\&$S performances, and finds a series of solutions by balancing two or more objectives need to be optimized. These solutions form a Pareto front that illustrates the trade-offs among different objectives, and provides designers with a range of choices to accommodate requirements of different applications and scenarios. \par
\vspace{1mm}
To achieve lower sensing error with higher communication sum rate, and study the trade-off C$\&$S performances in the ISAC system, an appropriate transmit waveform has to be designed by the following two optimization operations: \emph{i)} maximizing the communication sum rate while ensuring the SINR level required for the downlink communication users; \emph{ii)} minimizing the sensing error under the constraint of the transmit power buget for multiple sensing targets. Based on the formulated SOOPs in Sections \ref{2.a} and \ref{2.b}, we propose the MOOP, which is given by 
\vspace{1mm}
\begin{subequations}
\begin{align}
\mathrm{MOOP}:\quad\underset{\mathbf{x},\eta,\mathbf{R}}{\min}\quad
	&\mathbf{f}=\left[f_1(\mathbf{x}),f_2(\eta,\mathbf{R})\right]^T\label{17a}\\
	\mathrm{s.t.}\quad
	&\left|\mathrm{Im}\left(\mathbf{h}_k^H\mathbf{x}s_k^*\right)\right| \nonumber\\
	\le \mathrm{Re}\left(\mathbf{h}_k^H\mathbf{x}s_k^*\right)
	&\tan\phi-\sqrt{N_0\Gamma_k}\tan\phi,\forall{k} \in \mathcal{K},\label{19b}\\
	&\mathrm{Tr}(\mathbf{R}){\le}P_{\max},\label{19c}\\
		&\mathbf{R}\succeq 0 \label{19d},\\
	&\mathrm{rank}\{\mathbf{R}\}=1,\label{19e} 
\end{align}
\label{17}%
\end{subequations}
\vspace{1mm}
where $f_1(\mathbf{x})=-R_\mathrm{sum}$, $f_2(\mathbf{\eta,R})=M_s\left(\eta,\mathbf{R}\right)$.\par
\vspace{1mm}
The MOOP \eqref{17} is non-convex due to coupled sub-objective functions $f_1(\mathbf{x})$ and $f_2(\mathbf{R})$, as well as the non-convex quadratic term and log-sum operation of them. Note that the optimal solution of the objective function \eqref{17a} is affected by the weight of $f_1(\mathbf{x})$ and $f_2(\mathbf{R})$. We cannot solve MOOP \eqref{17} directly since the weight of $f_1(\mathbf{x})$ and $f_2(\mathbf{R})$ are unknown.\par

\section{Proposed Optimization Algorithm}\label{4}
In this section, we propose a weighted Tchebycheff-based transformation method to re-frame the above MOOP as a Pareto-optimal set solution problem. {This method considers the fact that equal resources bring different gains to C$\&$S performances in ISAC system, and aims to design optimal transmit waveforms for different performance preference settings by trading off the communication sum rate and radar sensing error.} Solving by the proposed iterative algorithms, the Pareto front can be obtain, which provides the insights for a flexible waveform design strategy under difference C$\&$S performance preferences in multi-user and multi-target ISAC scenarios.\vspace{-5pt}
\subsection{Trade-off of Dual-Fold Performance}\label{4.a}
	For MOOP \eqref{17}, the communication sum rate and the sensing MSE are coupled and conflicting that means improving one objective function inevitably leads to the deterioration of the other objective function. Specifically, a solution may be optimal for one objective while being suboptimal for the other objective. To address this issue, it is necessary to trade off  C$\&$S performances in multi-user and multi-target ISAC system.\par
{Pareto front composed of Pareto optimal solutions can trade off multiple objectives, where each solution represents an optimal design with different weights\cite{9022866,marler2004survey}. In this paper, we define the Pareto optimal transmit wavefrom as Definition \ref{def1}. A common approach for obtaining the Pareto front of the MOOP is to use scalarization method to transform multiple optimization objectives into a single objective by operation of weighted-sum\cite{9022866}. However, the weighted sum method is unfair, which ignores the fact that equal resources bring different gains to C$\&$S performances in ISAC system.} 

{ 
\begin{define}
 	The transmit wavefrom designs $\mathbf{x}^*$ are Pareto optimal if and only if there does not exist another point $\mathbf{x}'$ satisfying $f_i(\mathbf{x}'){\le}f_i(\mathbf{x}^*), \forall{i}{\in}\mathcal{P}$ and $f_j(\mathbf{x}')<f_j(\mathbf{x}^*), \exists{j}{\in}\mathcal{P}$ simultaneously, where $\mathcal{P}$ denotes the set of all SOOPs.\label{def1}
 \end{define} }

{In this subsection, we consider the gains of C$\&$S performances in ISAC system, and propose a weighted Tchebycheff-based transformation method to reformulate the MOOP into a Pareto optimal set solution problem, which introduces additional augmentation terms to avoid the presence of weakly Pareto optimal solutions in the Pareto front. Specifically, we optimize the proportional term of $f_i$ and $f_i^*$ rather than $f_i$ from a gain perspective, and the augmentation terms further minimize the scalarized global objective function which facilitate the acquisition of the Pareto optimal solution set. } 
{The $i$-th reformulated single objective function $f_i^\prime$ is
\begin{align}
f_i^\prime=\omega_i\left[\frac{f_i-f_i^*}{f_i^*}+\xi\sum_{j\in\mathcal{P}}\left(\frac{f_j-f_j^*}{f_j^*}\right)\right], \forall{i}{\in}\mathcal{P},\label{20}
\end{align}}
{where $\omega_i$ is the weight coefficient for the single optimization objective that represents the performance preferences of the decision-maker for the $i$-th SOOP and satisfies $\sum_{i\in\mathcal{P}}\omega_i=1$. Besides, a sufficiently small augmentation coefficient $\xi\in[0.001,0.01]$ is adopted to guarantee the Pareto optimality of the $i$-th SOOP. }\par
 By the weighted Tchebycheff-based transformation method, the MOOP \eqref{17} is reformulated as 
 \begin{subequations} 
 	\begin{align} 
 		\underset{\mathbf{x},\eta,\mathbf{R}}{\min}\quad&\underset{i\in\mathcal{P}}{\max}\left\{\omega_i\left[\frac{f_i-f_i^*}{f_i^*}+\xi\sum_{j\in\mathcal{P}}\left(\frac{f_j-f_j^*}{f_j^*}\right)\right]\right\}\label{21a}\\
 		\mathrm{s.t.}\quad  & \eqref{19b}, \eqref{19c}, \eqref{19d}, \eqref{19e}.
 	\end{align} \label{21}%
 \end{subequations}\par\vspace{-13pt}
In the problem \eqref{21}, $f_1$ and $f_2$ are abbreviated versions of $f_1(\mathbf{x})$ and $f_2( \eta,\mathbf{R})$, as well as $f_1^*$ and $f_2^*$ are their optimal solutions, respectively. Therefore, we have to prioritize the proposed SOOP1 and SOOP2 before solving the problem \eqref{21}. Next, for these non-convex SOOPs and problem \eqref{21}, we propose a series of iterative algorithms to design Pareto optimal transmit waveforms. The detailed iterative optimization process is described below.\vspace{-10pt}
\subsection{Optimization of Communication Sum Rate}
To obtain the optimal communication rum rate $f_1^*$, we need to solve SOOP1. However, it is noted that problem \eqref{SOOP1} is non-convex due to the log-sum operation in the objective \eqref{SOOP1a}. In addition, the joint optimization of the real and imaginary parts in constraint \eqref{SOOPb} increases the computational complexity. To address this issue, we develop the GP-based algorithm to extract a optimal solution in this subsection. 

Firstly, we define the channel of the rotated version as $\widetilde{\mathbf{h}}_k = \mathbf{h}_ks_k$ , and separate the real and imaginary parts of the complex vector as
\begin{align}
	\widetilde{\mathbf{h}}_k=\widetilde{\mathbf{h}}^R_k+j\widetilde{\mathbf{h}}_k^I,\\
	\mathbf{x}=\mathbf{x}^R+j\mathbf{x}^I,
\end{align}
where
$\widetilde{\mathbf{h}}_k^R=\mathrm{Re}\left(\widetilde{\mathbf{h}}_k\right), \widetilde{\mathbf{h}}_k^I=\mathrm{Im}\left(\widetilde{\mathbf{h}}_k\right), \mathbf{x}^R=\mathrm{Re}\left(\mathbf{x}\right), \mathbf{x}^I=\mathrm{Im}\left(\mathbf{x}\right); j=\sqrt{-1}$; and the useful signal received by the $k$-th user can be expressed as 
\begin{align}
	\widetilde{\mathbf{h}}_k^H\mathbf{x}
	&=\left(\widetilde{\mathbf{h}}^R_k\right)^T\mathbf{x}^R-\left(\widetilde{\mathbf{h}}^I_k\right)^T\mathbf{x}^I\nonumber\\
	&+j\left(\left(\widetilde{\mathbf{h}}^R_k\right)^T\mathbf{x}^I+\left(\widetilde{\mathbf{h}}^I_k\right)^T\mathbf{x}^R\right).
\end{align}

To convert the complex vector to a real-valued vector, we further define the real-valued vector as
\begin{align}
	\mathbf{z_k}=\left[\widetilde{\mathbf{h}}_k^R;\widetilde{\mathbf{h}}_k^I\right],\\ 
	\widetilde{\mathbf{x}}=\left[\mathbf{x}^I;\mathbf{x}^R\right].
\end{align}
Then, the real and imaginary parts of the complex vector $\widetilde{\mathbf{h}}_k^H\mathbf{x}$ are given by
\begin{align}
	&\mathrm{Im}\left(\widetilde{\mathbf{h}}_k^H\mathbf{x}\right)=\mathbf{z}_k^T\widetilde{\mathbf{x}},\\ &\mathrm{Re}\left(\widetilde{\mathbf{h}}_k^H\mathbf{x}\right)=\mathbf{z}_k^T\left[\mathbf{0,I;-I,0}\right]\widetilde{\mathbf{x}}=\widetilde{\mathbf{z}}_k^T\widetilde{\mathbf{x}},
\end{align}
where $\widetilde{\mathbf{z}}_k=[\mathbf{0,I;-I,0}]\mathbf{z}_k$. Thus, we rewrite receive signal  useful power of the $k$-th user as
\begin{align}
   \left|\widetilde{\mathbf{h}}_k^H\mathbf{x}\right|^2 =\left|\left|
   \left[
   \begin{matrix}
  (\widetilde{\mathbf{h}}_k^I)^T & \mathbf{0} \\
  \mathbf{0} & (\widetilde{\mathbf{h}}_k^R)^T \\
   \end{matrix} 
   \right] 
\left[ 
\begin{matrix}
	\mathbf{x}^I\\
	\mathbf{x}^R\\
\end{matrix}
\right]
\right|\right|^2=||\mathbf{H}^T_k\widetilde{\mathbf{x}}||^2.	
\end{align}%
After the above conversion, SOOP1 is rewritten as
\begin{subequations}
\begin{align} 
	\underset{\widetilde{\mathbf {x}}}{\max}\quad     
	& R_{\mathrm{sum}}= \sum^K_{k=1}B\log_2\left(1+\frac{||\mathbf{H}^T_k\widetilde{\mathbf{x}}||^2}{N_0}\right)\label{SOOP1.2a}\\     
	\mathrm{s.t.}\quad    	 
    	&\mathbf{z}_k^T\widetilde{\mathbf{x}}-\widetilde{\mathbf{z}}_k^T\widetilde{\mathbf{x}}\tan\phi+\sqrt{N_0\Gamma_k}\tan\phi \le0,\forall{k} \in \mathcal{K},\\
	    &-\mathbf{z}_k^T\widetilde{\mathbf{x}}-\widetilde{\mathbf{z}}_k^T\widetilde{\mathbf{x}}\tan\phi+\sqrt{N_0\Gamma_k}\tan\phi \le0,\forall{k} \in \mathcal{K},\\
	    &||\widetilde{\mathbf{x}}||^2 - P_{\max} \le 0.
\end{align} 
\label{SOOP1.2}%
\end{subequations}
Problem \eqref{SOOP1.2} is still non-convex since the objective function \eqref{SOOP1.2a} is a sum-log function. Fortunately, both the objective function and inequality constraints in problem \eqref{SOOP1.2} are monotonic, which means that problem \eqref{SOOP1.2} belong to the class of geometric programming (GP) problems. We can transform it into a convex optimization problem through variable substitution and function transformation by Lemma \ref{lemma2}. To this end, we transform the problem \eqref{SOOP1.2} into a convex problem by replacing the objective function $R_{sum}$ with $R_t(\mu_k)=\sum_{k=1}^K\log_2e^{\mu_k}$ and the variable $\mu_k$ that satisfies 
\begin{align}
	e^{\mu_k} = \frac{||\mathbf{H}^T_k\widetilde{\mathbf{x}}||^2 + N_0}{N_0}.
\end{align}
\begin{lemm}\label{lemma2}
	If we define GP problem as 
	\begin{subequations}
	\begin{align}
	\underset{x}{\max}\quad     
	& c_0x_1^{a_1}x_2^{a_2}...x_k^{a_k}, k\in \left\{1,2,\cdot\cdot\cdot,n\right\},\label{casa}\\     
	\mathrm{s.t.}\quad    	 
	&c_ix_{i,1}^{a_{i,1}}x_{i,2}^{a_{i,2}}...x_{i,k}^{a_{i,k}}\ge0, i=1,2,\cdot\cdot\cdot,m,\label{casb}
	\end{align}
\end{subequations}
where objective function \eqref{casa} and constraints \eqref{casb} are monotonic monomial, then the GP problem can be transformed into a convex optimization problem through variable substitution and function transformation.
\end{lemm}%
\begin{IEEEproof}
	See Appendix \ref{appendixA}.
\end{IEEEproof}
Then, we have 
\begin{subequations}
	\begin{align} 
		\underset{\widetilde{\mathbf {x}},{\mu_k}}{\max}\quad     
		&\sum_{k=1}^K\log_2e^{\mu_k}\label{f1a}\\     
		\mathrm{s.t.}\quad  
		&\frac{||\mathbf{H}^T_k\widetilde{\mathbf{x}}||^2 + N_0}{N_0} \ge e^{\mu_k}, \forall{k}\in \mathcal{K}\label{f1b},\\
		&\mathbf{z}_k^T\widetilde{\mathbf{x}}-\widetilde{\mathbf{z}}_k^T\widetilde{\mathbf{x}}\tan\phi+\sqrt{N_0\Gamma_k}\tan\phi \le0,\forall{k} \in \mathcal{K}\label{f1c},\\
		&-\mathbf{z}_k^T\widetilde{\mathbf{x}}-\widetilde{\mathbf{z}}_k^T\widetilde{\mathbf{x}}\tan\phi+\sqrt{N_0\Gamma_k}\tan\phi \le0,\forall{k} \in \mathcal{K}\label{f1d}, \\
		&||\widetilde{\mathbf{x}}||^2 - P_{\max} \le 0.\label{f1e}
	\end{align} 
\label{33}%
\end{subequations}%
\begin{algorithm}[t]
	\caption{ GP-Based Algorithm for Solving SOOP1}
	\renewcommand{\algorithmicrequire}{\textbf{Input:}}
	\renewcommand{\algorithmicensure}{\textbf{Output:}}
	\begin{algorithmic}[1]
		\REQUIRE Initialized $\mathbf{x}^{(0)},\mu_k^{(0)}$, the tolerance $\epsilon$, the  iteration number $t_1=0$.
		\REPEAT
		\STATE Compute $\widetilde{\mathbf{x}}^{(t_1)}=[\mathrm{real}({\mathbf{x}^{(t_1)}}),\mathrm{imag}({\mathbf{x}^{(t_1)}})]^T$;
		\STATE Compute $\mu_k^{(t_1)}=\frac{||\mathbf{H}^T_k\widetilde{\mathbf{x}}^{(t_1)}||^2 + \sigma^2}{N_0 \log_2e}, k \in \mathcal{K}$;
		\STATE Obtain $\mathbf{x}^{(t_1+1)}$ by solving the GP problem \eqref{33};
		\STATE Update: $t_1=t_1+1$;
		\UNTIL{$|\sum_{k=1}^K{\mu_k}^{(t_1)}-\sum_{k=1}^K{\mu_k}^{(t_1-1)}|\le\epsilon_1$;} 
		\ENSURE  The  optimal transmit waveform $\mathbf{x}^*$.
	\end{algorithmic}
	\label{algorithm1}
\end{algorithm}%
We observe that 
\begin{align}
	\sum_{k=1}^K\log_2e^{\mu_k}=\log_2e\sum_{k=1}^K{\mu_k}.
\end{align}
Thus, problem \eqref{33} is a convex optimization problem. The GP-based transmit waveform design algorithm is summarized in Algorithm \ref{algorithm1}, and it is convergent due to objective function \eqref{SOOP1.2a} is monotonic proven in Appendix \ref{appendixA}.
\vspace{-3mm}
\subsection{Optimization of Sensing Error}
{In this subsection, we intend to solve the non-convex SOOP2 for obtaining the minimum radar sensing MSE $f_2^*$. It is challenging to obtain the variable $\mathbf{R}$ due to the non-convex rank-1 constraint \eqref{16d}. We first derive the closed-form optimum solution of $\eta$ and employ the semidefinite relaxation technique to drop the rank-1 constraint.} \par
{According to SOOP2, it is evident that the variable $\eta$ only appears in the objective function \eqref{16a}. As the objective function \eqref{16a} is a quadratic convex function, we can derive the closed-form optimum solution of $\eta$ by applying the first-order optimal conditions. The objective function \eqref{16a} attains the minimum value when its gradient equals zero, i.e.,
	\begin{align}
		\frac{\partial M_s\left(\eta,\mathbf{R}\right)}{\partial \eta} &= \frac{1}{L}\sum\nolimits^L_{l=1}\left(2\eta\hat{G}^2(\theta_l)-2\hat{G}(\theta_l)\mathbf{a}(\theta_l)^H\mathbf{R}\mathbf{a}(\theta_l)\right)\nonumber\\
		&= 0. \label{34}
	\end{align}
Eqn. \eqref{34} gives rise to the closed-form optimum solution of $\eta$, as given by 
	\begin{align}
		\eta^*&=\frac{\sum_{l=1}^{L}\hat{G}^2(\theta_l)\mathbf{a}^H(\theta_l)\mathbf{R}\mathbf{a}(\theta_l)}{\sum_{l=1}^{L}\hat{G}^2(\theta_l)}\nonumber\\
		&=\frac{\sum_{l=1}^{L}\hat{G}^2(\theta_l)\mathrm{vec}(\mathbf{A}_l)^H\mathrm{vec}(\mathbf{R})}{\sum_{l=1}^{L}\hat{G}^2(\theta_l)},\label{eta} 
	\end{align}	
	where $\mathbf{A}_l=\mathbf{a}(\theta_l)\mathbf{a}^H(\theta_l)$, $\mathrm{vec}(\mathbf{R})$ and $\mathrm{vec}(\mathbf{A}_l)$ denote the vectorization of $\mathbf{R}$ and $\mathrm{vec}(\mathbf{A}_l)$, respectively. By substituting \eqref{eta} into \eqref{16a}, the objective function can be re-expressed as 
	\begin{align}
		M_s(\mathbf{R})=\sum_{l=1}^L\left|\hat{G}(\theta_l)\mathbf{g}_l^H\mathrm{vec}(\mathbf{R}) \right|^2\label{M_s},
	\end{align}
	where $\mathbf{g}_l^H=\frac{\sum_{l=1}^L\hat{G}(\theta_l)\mathrm{vec}(\mathbf{R})}{\sum_{l=1}^L\hat{G}^2(\theta_l)}-\mathrm{vec}(\mathbf{A}_l)^H$.}\par
 {Next, considering that the non-convexity in SOOP2 is primarily attributed to the rank-1 constraint \eqref{16d}, the SDR method is employed to drop this constraint. By replacing \eqref{14} with \eqref{M_s}, the relaxed SOOP2 can be reformulated as 
\begin{subequations}\label{38}
		\begin{align}
			&\underset{\mathbf{R}}{\min}\quad M_s(\mathbf{R}) \\
			&	\mathrm{s.t.}\quad \eqref{16b} , \eqref{16c}.
		\end{align}\label{38}%
\end{subequations}}
{It can be observed that the relaxed form of problem \eqref{38} is a convex Quadratic semidefinite programming problem (QSDP). As a result, the optimal solution $\mathbf{R}^*$ can be efficiently solved by using widely-used toolboxes such as CVX \cite{grant2014cvx}. It is noted that the SDR approach generates an approximated solution to the optimization problem \eqref{38} by neglecting the rank-1 constraint \eqref{16d}, which may result in a solution with high rank. The Gaussian randomization method \cite{5447068} can be employed to generate a sub-optimal solution $\mathbf{x}^*$ for problem \eqref{38} when the obtained solution $\mathbf{R}^*$ does not satisfy the rank-1 condition.} \par
\begin{algorithm}[t]
	\caption{ SDR-Based Algorithm for Solving SOOP2}
	\renewcommand{\algorithmicrequire}{\textbf{Input:}}
	\renewcommand{\algorithmicensure}{\textbf{Output:}}
	\begin{algorithmic}[1]
		\REQUIRE Initialized $\mathbf{R}^{(0)}$, the tolerance $\epsilon_2$, the iteration number  $t_2=0$.
		\REPEAT 
		\STATE Compute $\eta^{(t_2)}\frac{\sum_{l=1}^{L}\hat{G}^2(\theta_l)\mathrm{vec}(\mathbf{A}_l)^H\mathrm{vec}(\mathbf{R}^{(t_2)})}{\sum_{l=1}^{L}\hat{G}^2(\theta_l)}$;
		\STATE Compute $\mathbf{R}^{(t_2)}$ by solving the QSDP \eqref{38} with CVX;
		\STATE Compute $M_s\left(\eta^{(t_2)},\mathbf{R}^{(t_2)}\right)$;
		\STATE Update $t_2:=t_2+1$;
		\UNTIL{$|M_s\left(\eta^{(t_2)},\mathbf{R}^{(t_2)}\right)-M_s\left(\eta^{(t_2-1)},\mathbf{R}^{(t_2-1)}\right)|\le\epsilon_2$;} 
		\STATE Obtain $\mathbf{x}^*$ by eigenvalue decomposition of $\mathbf{R}^*$;
		\ENSURE The optimal transmit waveform $\mathbf{x}^*$.
	\end{algorithmic}
	\label{algorithm2}
\end{algorithm}%
{The SDR-based algorithm for solving SOOP2 is summarized in Algorithm \ref{algorithm2}. The primary computational challenge in Algorithm \ref{algorithm2} arises from solving the QSDP \eqref{38}. For a given solution accuracy $\epsilon_2$, the complexity of Algorithm \ref{algorithm2} is mainly determined by the process of solving QSDP \eqref{38}, which is given by $\mathcal{O}(LNN_T^4\log(1/\epsilon_2))$.}
\vspace{-2mm}
\subsection{Optimization of Integrated Waveform}\label{4-d}

In this subsection, by using the optimal solutions of SOOP1 and SOOP2, we propose an SCA-based iterative optimization algorithm to solve the problem \eqref{21}. It is important to  note that the convexity of the objective function \eqref{21a} is only determined by $f_1$ and $f_2$ due to the fact that equation \eqref{20} is linear and the $\max\{\cdot\}$ operator is convex-preserving, which means we can confirm the objective function \eqref{21a} is convex as long as ensure that $f_1$ and $f_2$ are convex. Therefore, for a more intuitive solution process, we will directly analyze MOOP \eqref{17} and then reformulate it by the proposed weighted Tchebycheff-based transformation method in Section \ref{4.a}.

The MOOP \eqref{17} is still difficult to solve optimally. On the one hand, variables $\mathbf{x}$ and $\mathbf{R}$ to be optimized are coupled to each other in the MOOP \eqref{17}, and it generally belongs to NP-hard; On the other hand, objective function \eqref{17a} is non-convex due to the operation on the sum of logarithms in $f_1(\mathbf{x})$ and the presence of rank-1 constraint \eqref{19e}. Regarding the issue above, we introduce an auxiliary variable $c$ and a new constraint $(\mathbf{h}_k^H\mathbf{x}s_k^*)^2 \le c^2$ to reformulate problem \eqref{17}. We can see from \eqref{7} that variable $\mathbf{x}$ and variable $\mathbf{R}$ satisfy the relationship of $\mathbf{R}=\mathbf{x}\mathbf{x}^H$. Therefore, we have $\mathbf{h}_k^H\mathbf{R}\mathbf{h}_k \le c^2$. Then, we replace the polynomial $\frac{||\mathbf{h}^T_k\mathbf{x}||^2 + N_0}{N_0}$ with variable $e^{\mu_k}$ , which also generates a new constraint as $e^{\mu_k} \le \frac{c^2 + N_0}{N_0}$. Then, we reformulate the MOOP as
\begin{subequations}
	\begin{align}
		\underset{\mathbf{R},c,\mu_k}{\min}\quad
		&\mathbf{f}=\left[-R_t(\mu_k),M_s(\mathbf{R})\right]^T\label{f3.3a}\\
		\mathrm{s.t.}\quad
			&\frac{c^2 + N_0}{N_0} \ge e^{\mu_k}, \forall{k} \in \mathcal{K},\label{f3.3b}\\
		&\left|\mathrm{Im}(c)\right| \le \left(\mathrm{Re}(c)-\sqrt{N_0\Gamma_k}\right)\tan\phi,\forall{k} \in \mathcal{K},\label{f3.3c}\\
		&\mathbf{h}_k^H\mathbf{R}\mathbf{h}_k \le c^2,   \label{f3.3d}\\   
			&\eqref{19c}, \eqref{19d}, \eqref{19e}.
	\end{align}
\label{f3.3}%
\end{subequations}
However, problem \eqref{f3.3} remains a non-convex problem due to the quadratic terms in the constraints \eqref{f3.3b} and \eqref{f3.3d}. we develop the SCA-based algorithm to solve the problem \eqref{f3.3}. The first-order Taylor expansion of $c^2$ at the point $\overline{c}$ can be given by
\begin{align}
c^2=\overline{c}^2+2\overline{c}(c-\overline{c}), \label{1}
\end{align}
where $\overline{c}$ is feasible to the problem \eqref{f3.3}. The constraints \eqref{f3.3b} and \eqref{f3.3d} in problem \eqref{f3.3} can be rewritten as 
\begin{align}
	\frac{\overline{c}^2+2\overline{c}\left(c-\overline{c}\right) + N_0}{N_0} \ge e^{\mu_k}, \forall{k} \in \mathcal{K},\label{1.1}
\end{align}
\begin{align}
	\mathbf{h}_k^H\mathbf{R}\mathbf{h}_k \le \overline{c}^2+2\overline{c}\left(c-\overline{c}\right)\label{1.2}.
\end{align}

It is observed that \eqref{1.1} and \eqref{1.2} are convex since \eqref{1} is linear and convex. Then, by replacing \eqref{f3.3b} and \eqref{f3.3d} with \eqref{1.1} and \eqref{1.2}, we have 
\begin{subequations}
		\begin{align} 
		\underset{\mathbf{R},c,\mu_k,y_l}{\min}\quad     
		&\mathbf{f}=\left[-R_t(\mu_k),M_s(\mathbf{R})\right]^T \label{41a} \\ 
		\mathrm{s.t.}\quad 
		&\frac{\overline{c}^2+2\overline{c}\left(c-\overline{c}\right) + N_0}{N_0} \ge e^{\mu_k}, \forall{k} \in \mathcal{K},\label{43b}\\    
		&\left|\mathrm{Im}(c)\right| \le \left(\mathrm{Re}(c)-\sqrt{N_0\Gamma_k}\right)\tan\phi,\forall{k} \in \mathcal{K},\label{43c}\\ 
		&\mathbf{h}_k^H\mathbf{R}\mathbf{h}_k \le \overline{c}^2+2\overline{c}\left(c-\overline{c}\right),   \label{43d}\\    
	&\eqref{19c}, \eqref{19d}, \eqref{19e}
		\end{align}
	\label{f3.4}%
\end{subequations}
\begin{algorithm}[t]
	\caption{SCA-Based Algorithm for Solving MOOP}
	\renewcommand{\algorithmicrequire}{\textbf{Input:}}
	\renewcommand{\algorithmicensure}{\textbf{Output:}}
	\begin{algorithmic}[1]
		\REQUIRE 
		The chosen preference coefficients $\omega_1^{(0)}$ and $\omega_2^{(0)}$;
		The preference coefficient step size $\Delta\omega$; The preference index $t=1$.
		\STATE Execute Algorithm \ref{algorithm1} to obtain the optimal $f_1^*$;
		\STATE Execute Algorithm \ref{algorithm2} to obtain the optimal $f_2^*$;
		\WHILE{$t < 1/{\Delta\omega}$}
		\STATE Initialize $\mathbf{R},c,\{y_l\}_{l=1}^L,t$ and the iteration index $t_3=0$;\\
		\REPEAT 
		\STATE Substitute $f_1^*$ and $f_2^*$, $\omega_1^{(t)}$, $\omega_2^{(t)}$ into problem \eqref{f3.6} to obtain $c^{(t_3)}$;
		\STATE Upadte ${\overline{c}}^{(t_3+1)} =c^{t_3}$;
		\STATE set $t_3=t_3+1$;
		\UNTIL{$|c^{(t_3)}-c^{(t_3-1)}|\le\epsilon_3$, where $\epsilon_3>0$ is a sufficiently small positive number.};
		\STATE Obtain $\mathbf{x}_t^*$ by decompose of $\mathbf{R}_t^*$ when the rank$(\mathbf{R}_t^*)=1$; otherwise the Gaussian Randomization method is utilized to obtain a rank-1 approximation;
		\STATE Set $t_3=0$ and Update $\omega_1^{(t)}=\omega_1^{(t)}+\Delta\omega$, $\omega_2^{(t)}=\omega_2^{(t)}-\Delta\omega$, and $t=t+1$;
		\ENDWHILE	
		\ENSURE  
		Pareto optimal waveforms set $\{\mathbf{x}_1^*,\mathbf{x}_2^*,...,\mathbf{x}_t^*\}$. 
	\end{algorithmic}
	\label{algorithm3}
\end{algorithm}
The functions $f_1$ and $f_2$ in \eqref{41a} are redefined as
\begin{align}
	&f_1=-R_t(\mu_k)=-\log_2e\sum_{k=1}^K e^{\mu_k},\label{44}\\
	&f_2=M_s(\mathbf{R})=\sum_{l=1}^L\left|\hat{G}(\theta_l)\mathbf{g}_l^H\mathrm{vec}(\mathbf{R}) \right|^2.\label{45}
\end{align}\par
By the weighted Tchebycheff-based transformation method, the corresponding MOOP \eqref{f3.4} can be rewritten as  \vspace{-0.3em}
 \begin{subequations}
 	\begin{align}
 	\underset{\mathbf{R},c,\mu_k,y_l}{\min}\quad&\alpha\\	
 	\mathrm{s.t.}\quad 
 	    &\omega_1\left(\frac{f_1-f_1^*}{f_1^*}+\xi\sum_{j=1}^{2}\left(\frac{f_j-f_j^*}{f_j^*}\right)\right)\le \alpha,\label{46b}\\
 	    &\omega_2\left(\frac{f_2-f_2^*}{f_2^*}+\xi\sum_{j=1}^{2}\left(\frac{f_j-f_j^*}{f_j^*}\right)\right) \le \alpha,\label{46c}\\
 	&\eqref{43b}, \eqref{43c}, \eqref{43d}, \eqref{19c}, \eqref{19d}, \eqref{19e}\label{46d}
	\end{align}
\label{f3.6}%
 \end{subequations} 
where $\alpha$ is an introduced auxiliary variable.
We can observe that the optimization problem \eqref{f3.6} without Rank-1 constraint \eqref{19e} is a standard convex problem and can be solving by using the well-known toolbox such as CVX \cite{grant2014cvx}. Thus, the Pareto front in terms of the sum MSE and achievable sum rate can be obtained by varying the values of $\omega_1 $ and $\omega_2$ (note that $\omega_1 + \omega_2=1$ holds). Note that the Gaussian randomization method \cite{5447068} can be employed to generate a sub-optimal solution $\mathbf{x}^*$ for problem \eqref{f3.6} if the obtained solution $\mathbf{R}^*$ fails to be rank-1.
For clarity, the proposed iterative algorithm for problem \eqref{21} is shown detailedly in Algorithm \ref{algorithm3}.
\begin{prop}\label{Prop1}
The problem in \eqref{17} converges when the SCA-based iterative algorithm is used.
\end{prop}
\begin{IEEEproof}
	See Appendix \ref{appendixC}.
\end{IEEEproof}
The complexity of Algorithm \ref{algorithm3} is not only caused by the number of Pareto optimal solutions in Pareto front, but also depends on the execution of Algorithm \ref{algorithm1} and Algorithm \ref{algorithm2}. The complexity of Algorithm \ref{algorithm1} is mainly determined by the matrix variables of size $N_t \times N_t$, which is $\mathcal{O}\left(K^4N_t^4+\left(4K+1\right)N_t^2\log(1/\epsilon_1)\right)$. The complexity of Algorithm \ref{algorithm2} is mainly determined by the process of solving QSDP \eqref{38}, which is $\mathcal{O}(LNN_T^4\log(1/\epsilon_2))$. Thus, the complexity of Algorithm \ref{algorithm3} is given by $1/{\Delta\omega}\cdot\mathcal{O}$$\left(LNN_T^4+K^4N_t^4+\left(4K+1\right)N_t^2\log(1/\epsilon_3)\right)$, which is substantially less than the exhaustive search method.
\section{Numerical Results}\label{5}
\begin{table*}[t] 
	\centering 
	\caption{Simulation Parameters}
	\begin{tabular}{|c|c||c|c|} \hline 
		Parameter & Value & Parameter & Value\\ \hline\hline
		Number of transmit antennas $N_t$ & 8 &Number of receive antennas $N_r$ &	8 \\ \hline
		Transmit power budget $P_{\max}$ & 25 $\mathrm{dBm}$&		The noise power at Users $N_0$& -60 dBm  \\ \hline
		Number of communication users $K$ & 3 &Number of sensing targets $N$ & 3  \\ \hline
		The performance preference step size $\Delta\omega$ & 0.01  &
		Number of sample angles $L$ & 180  \\ \hline
		The width of desired beampattern $\Delta\theta$ & $3^\circ$  &
		Rician factor $\nu_k$ & 1 \\ \hline
		The augmentation coefficient $\xi$ & 0.001  &
		Error tolerance $\epsilon_1$ & $10^{-4}$ \\ \hline
		Error tolerance $\epsilon_2$ & $10^{-4}$ &
		Error tolerance $\epsilon_3$ & $10^{-4}$ \\ \hline
	\end{tabular}
	\label{table1}
\end{table*}
Some numerical simulations are conducted to evaluate the performance of our proposed waveform approach in terms of the achievable trade-off, including the MOOP transformation method and optimization algorithms. Our test environment is MATLAB R2019b on a desktop computer with 3.20 GHz Intel Core i7-8700 CPU and 16GB random access memory (RAM). In the simulations, three sensing targets are placed at angles of $30^\circ$, $90^\circ$, and $150^\circ$ with respect to the BS, respectively. Unless otherwise specified, all other simulation parameters of the proposed ISAC system and related algorithms are listed in Table \ref{table1}. It is worth noting that the Pareto optimal solutions for different system settings are obtained by varying the value of $\omega_i$ with $\sum_{i\in \{1,2\}}\omega_i=1$, and all the simulation results are obtained through 1000 Monte Carlo simulations. \par
\begin{figure}[t]
	\centering{}\includegraphics[width=3.5in]{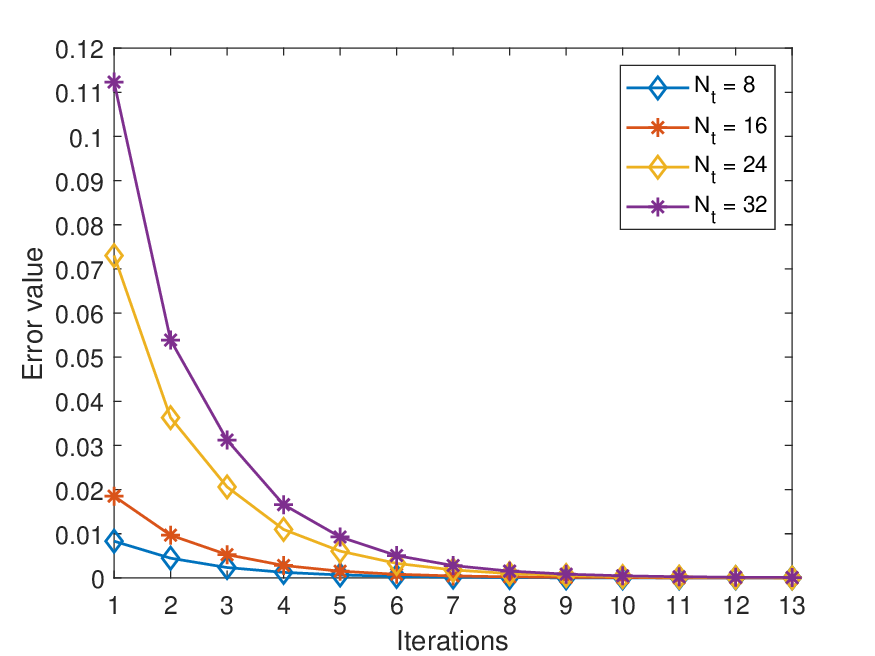}
	\caption{Convergence behavior of Algorithm 3.}
	\label{fig:iteration}
\end{figure}
Fig. \ref{fig:iteration} shows the convergence of the proposed Algorithm \ref{algorithm3} under the performance preference $\omega_1=\omega_2=0.5$. In each iteration, the error values between the current iteration and the previous iteration for different antenna settings are different, and it increases with as the number of transmit antennas. This is because the complexity of the proposed Algorithm \ref{algorithm3} is mainly influenced by the number of antennas as demonstrated in Section \ref{4-d}. We also observe that the proposed Algorithm \ref{algorithm3}  approximately converges for all settings after about 13 iterations.\par
\begin{figure}[t]
	\centering{}\includegraphics[width=3.5in]{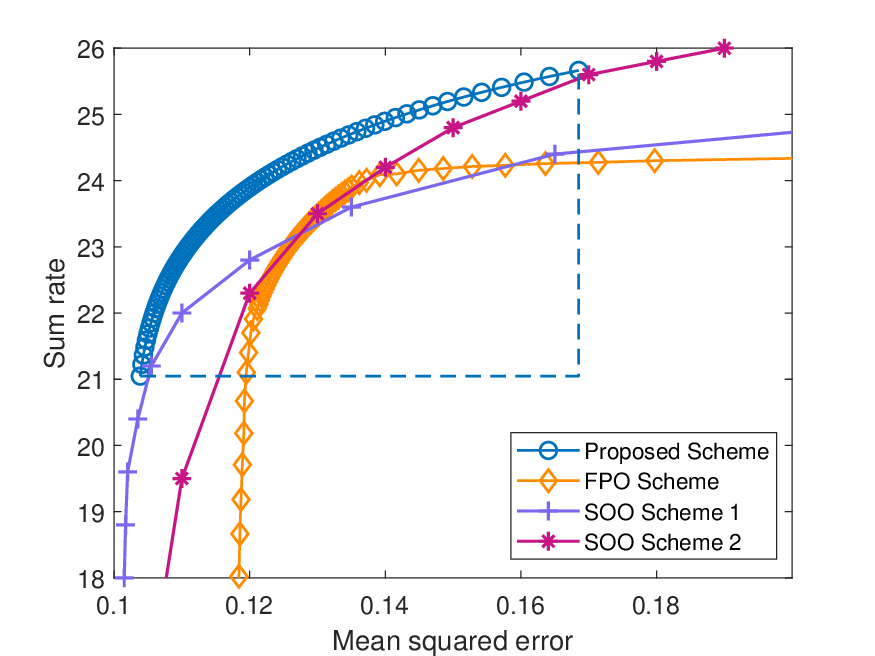}
	\caption{Trade-off between sum rate and sensing MSE achieved by different schemes.}
	\label{fig:comparison}
\end{figure}
\begin{figure*}[t]
	\subfigure[]{
		\centering{}\includegraphics[width=3.5in]{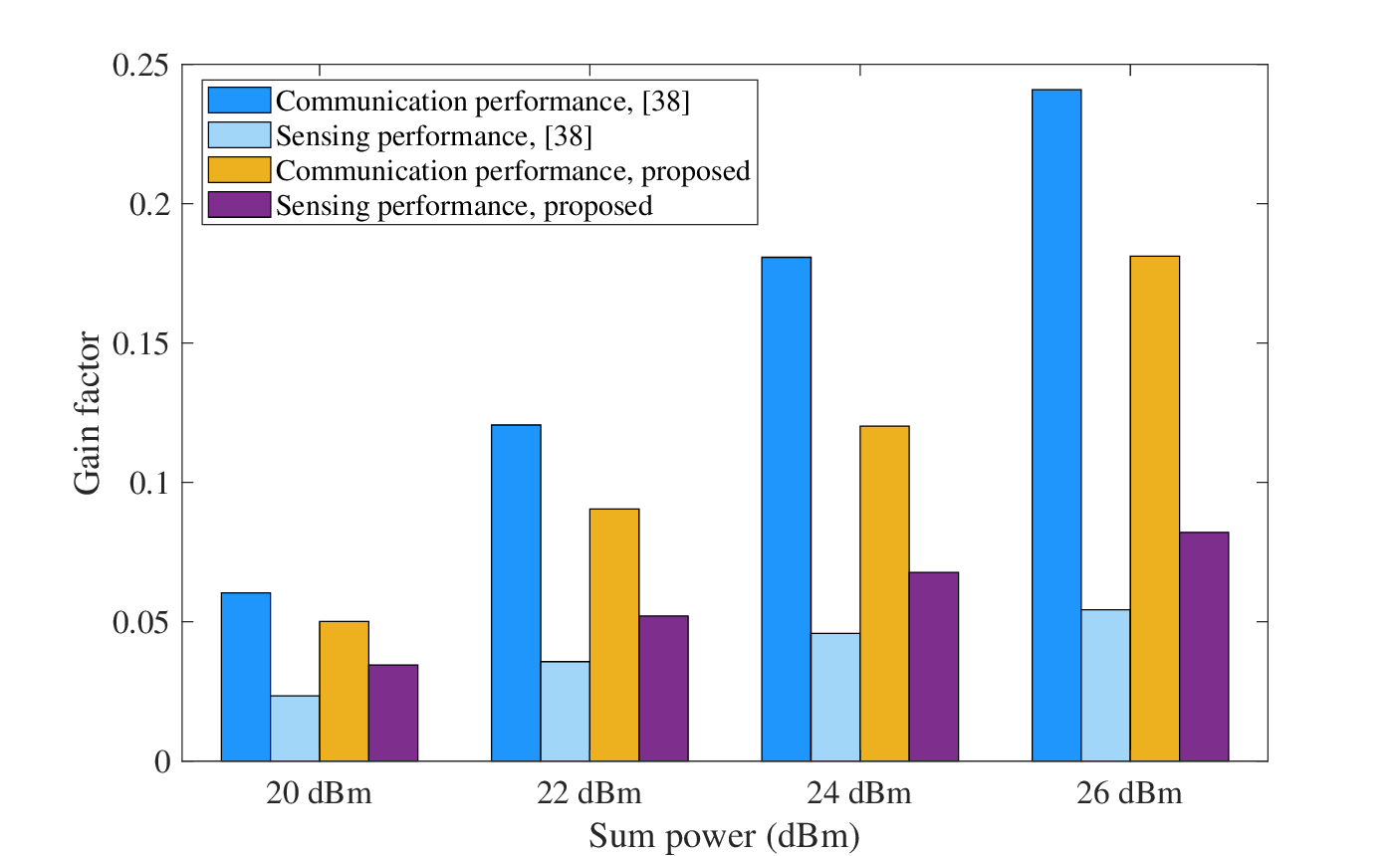}\label{fig:gain_power}
	}
	\subfigure[]{
		\centering{}\includegraphics[width=3.5in]{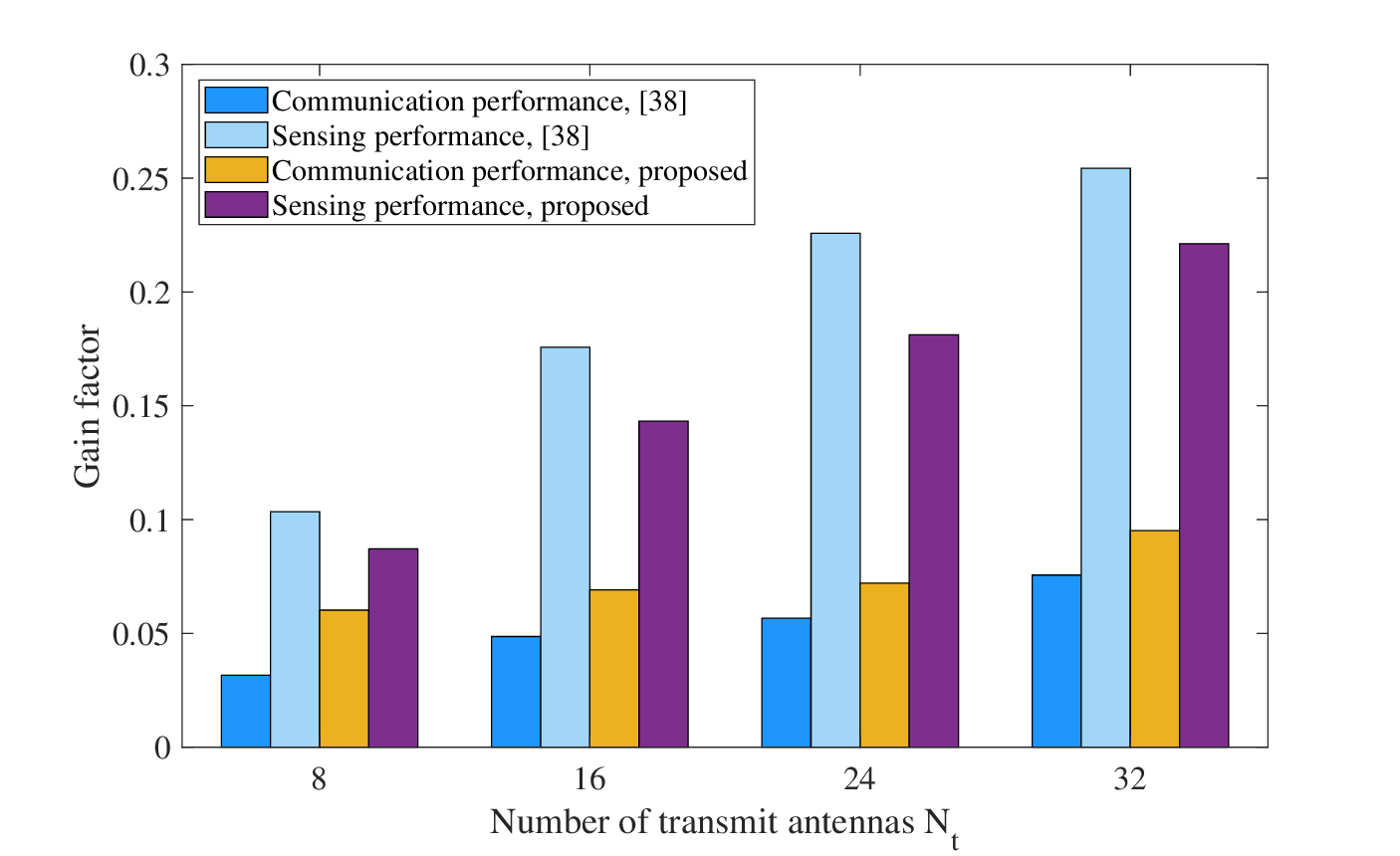}
		\label{fig:gain_antennas}
	}
	\caption{Performance gain of communication and sensing versus (a) sum power, and (b) number of transmit antennas }
	\label{fig:gain}
\end{figure*}
In Fig. \ref{fig:comparison}, we compare the performance limit of the proposed scheme with the fairness profile optimization (FPO) scheme in \cite{10007645}. Each point on the Pareto front of our proposed algorithm represents a optimal balance solution, x- and y-axes of which indicates the achievable sum rate and sensing MSE without sacrificing another performance, respectively. It is seen that the Pareto front obtained from the proposed scheme lies above the performance boundary of the benchmark scheme, which indicates that the proposed scheme achieves higher communication rate and lower sensing MSE compared to the FPO scheme. In addition, in order to highlight the advantages of the proposed MOO algorithm compared with single-objective optimization (SOO), Fig. 3 considers two SOO schemes for comparison: 
	\begin{itemize}
		\item \textbf{SOO scheme 1}: Sensing MSE is set as the objective subject to the communication rate constraint. To ensure a wide traversal range, the communication rate threshold of each user is within $4 \sim 8$ bps;
		\item \textbf{SOO scheme 2}: Communication rate is optimized subject to the sensing MSE constraint. To ensure a wide traversal range, the beampattern MSE threshold is within $0.1 \sim 0.2 $. 
	\end{itemize}
Note that the weights characterizing performance preferences are not employed in SOO schemes since specific constraint indicators are set. The performance boundaries of SOO schemes can be obtained by adjusting the threshold of the sensing/communication constraints \cite{9124713}.  Simulation results show that the proposed MOO scheme has better trade-off performance, while the SOO schemes only yield better single performance of communication or sensing.  \par
To prove the fairness of the proposed method in terms of performance gain, Fig. \ref{fig:gain} compares the proposed scheme with the weighted-sum method in \cite{9022866}. Under the same weight setting ($\omega_1=\omega_2=0.5$), the proposed method has a smaller gain gap in both C$\&$S performances compared to the weighted-sum method, confirming the fairness of the proposed method. Additionally, Fig. \ref{fig:gain} provides initial insights into the impact of transmitter parameters, namely, transmit power and the number of transmit antennas, on the ISAC system. Comparing Figs. 4(a) and 4(b), we observe that communication performance is mainly determined by the transmit power, while sensing performance is primarily influenced by the number of transmit antennas. To further validate this insight, we conducted the trade-off between C$\&$S performance in the following simulations.\par
\begin{figure}[t]
	\centering{}\includegraphics[width=3.5in]{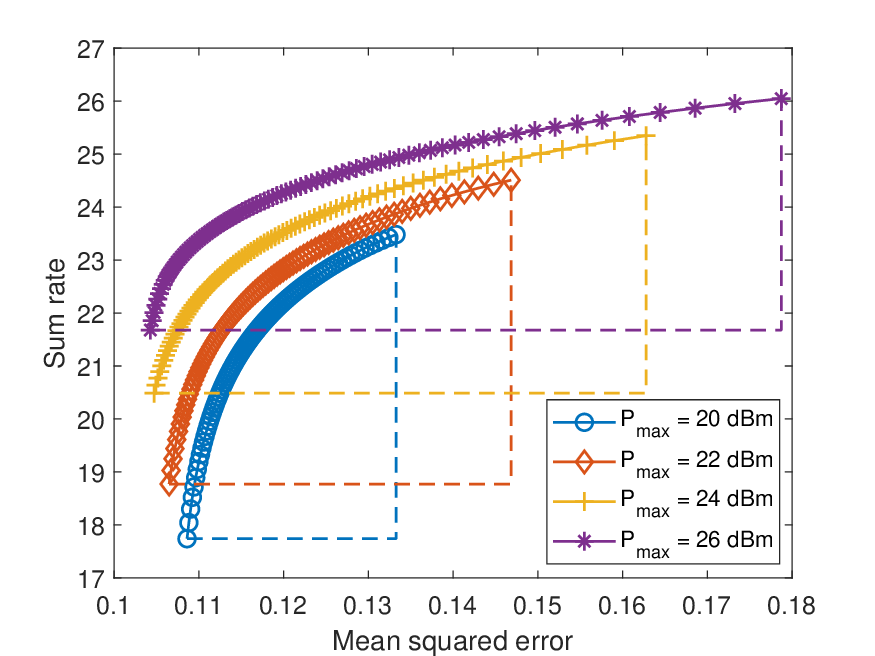}
	\caption{Trade-off between sum rate and sensing MSE under different power budget.}
	\label{fig:power trade-off}
\end{figure}
\begin{figure}[t]
	\centering{}\includegraphics[width=3.5in]{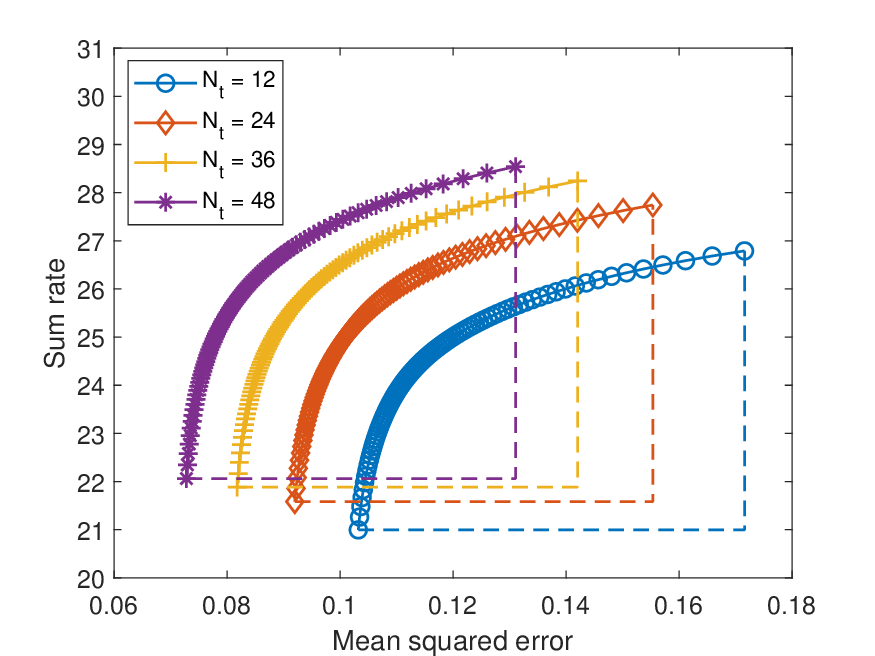}
	\caption{
		Trade-off between sum rate and sensing MSE under different number of transmit antennas.}
	\label{fig:antenna trade-off}
\end{figure}
\begin{figure}[t]
	\centering{}\includegraphics[width=3.5in]{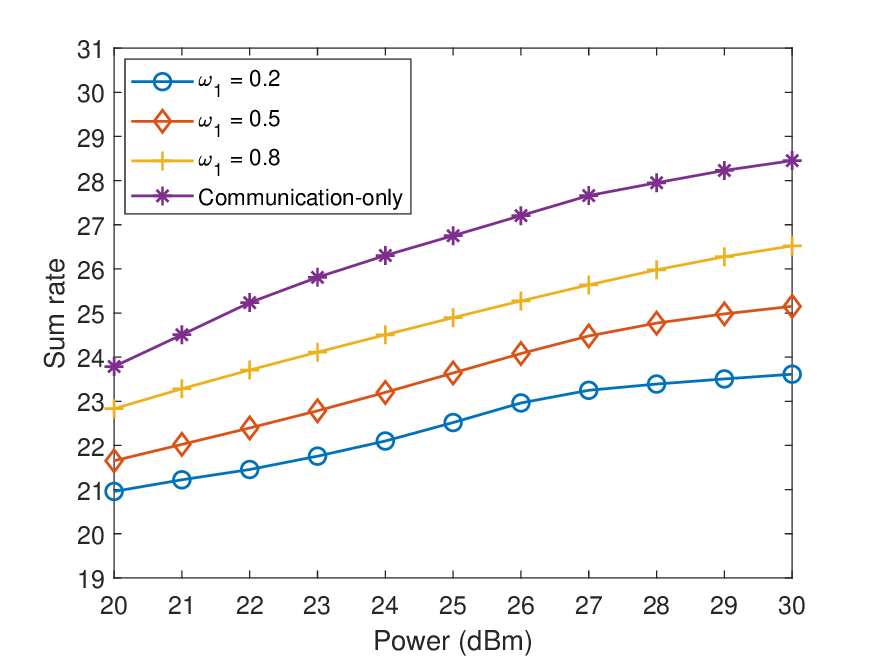}
	\caption{Communication sum rate versus transmit power budget under different communication preferences.}
	\label{fig:power communication}
\end{figure} 
\begin{figure}[t]
	\centering{}\includegraphics[width=3.5in]{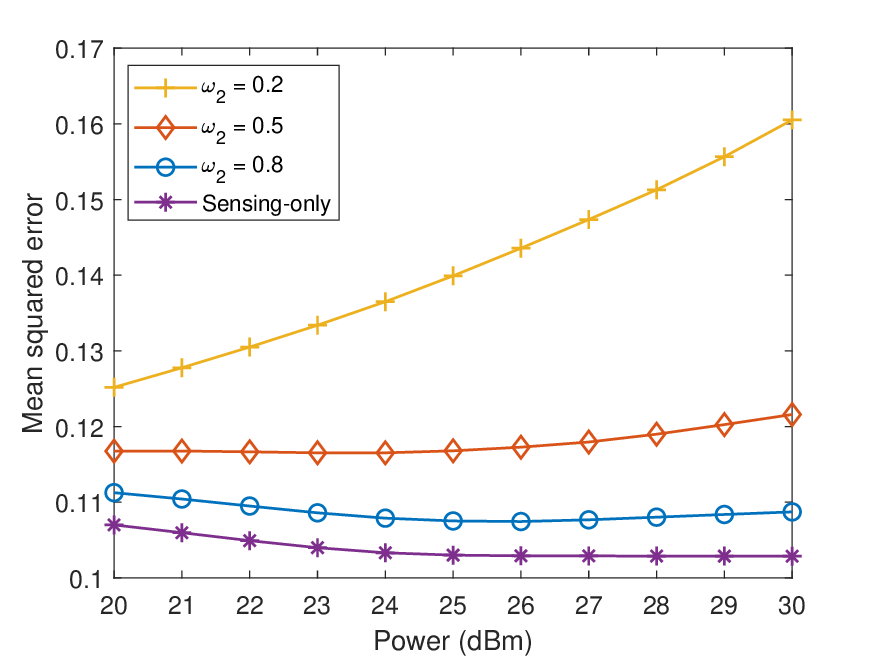}
	\caption{Sensing MSE versus transmit power budget under different sensing preferences.}
	\label{fig:power sensing}
\end{figure}
Figs. \ref{fig:power trade-off} and \ref{fig:antenna trade-off} present the trade-offs between achievable sum rate and sensing MSE of the proposed ISAC system under different settings of transmit power budget and number of transmit antennas, respectively. In both Fig. \ref{fig:power trade-off} and \ref{fig:antenna trade-off}, it can be  observed that the sensing MSE deteriorates as the achieve sum rate increases, and vice versa. This phenomenon confirms the conflict between C$\&$S performances in ISAC systems from a trade-off perspective. Interestingly, we also observe that the geometric shapes of the Pareto front and the Pareto feasible region under different settings  are various in Fig. \ref{fig:power trade-off} but remain constant in Fig. \ref{fig:antenna trade-off}. This indicates that the coupling between C$\&$S performances mainly occurs in the domain of transmit power budget rather than the number of transmit antennas. {The reason is that the increase of transmit antennas will bring the additional degrees of freedom to the ISAC system, which is beneficial to both C$\&$S performances. In addition, the increasing transmit power results in larger sidelobe gains in non-target directions, which is beneficial for communication but undesirable for sensing performance.}\par

 \begin{figure}[t]
	\centering{}\includegraphics[width=3.5in]{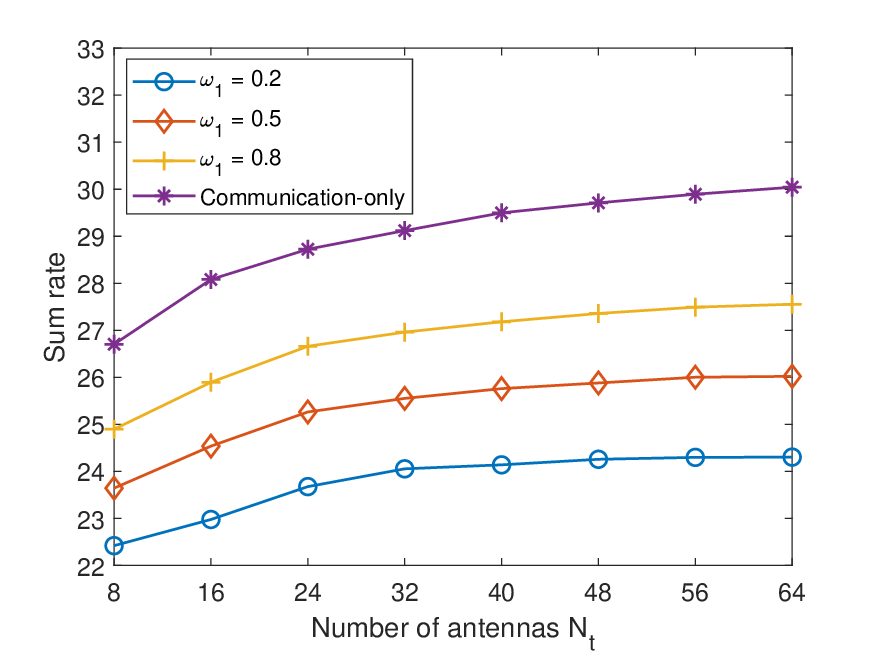}
	\caption{Communication sum rate versus number of transmit antennas under different communication preferences.}
	\label{fig:antenna communication}
\end{figure}
\begin{figure}[t]
	\centering{}\includegraphics[width=3.5in]{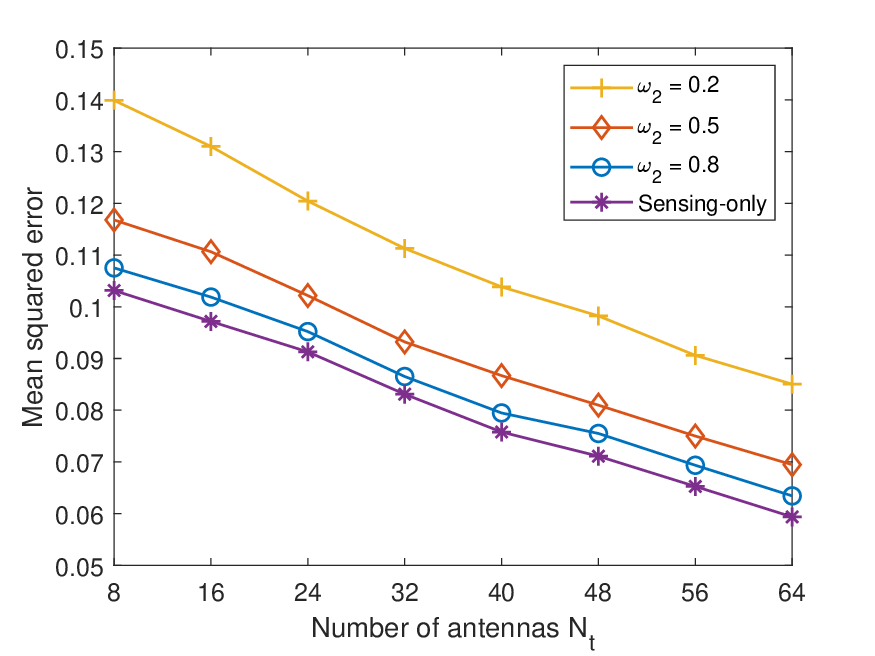 }
	\caption{Sensing MSE versus number of transmit antennas under different sensing preferences.}
	\label{fig:antenna sensing}
\end{figure}
In Figs. \ref{fig:power communication} and \ref{fig:power sensing}, we analyze the effect of performance preference on achievable sum rate and sensing MSE separately under different transmit power budget settings. In Fig. \ref{fig:power communication}, we observe that the  achieve sum rate monotonically increases with the transmit power budget for all performance preferences. Furthermore, the achievable sum rate increases by 3.5 bit/s with performance preference $\omega_1=0.8$ when the transmit power budget is upgraded from 20 $\mathrm{dBm}$ to 30 $\mathrm{dBm}$, but it only increases by 2.7 bit/s with performance preference $\omega_1=0.2$. It is evident that the achievable sum rate with higher communication performance preference increases more rapidly with transmit power budget due to more power is allocated to the communication in such settings. Fig. \ref{fig:power sensing} shows an interesting phenomenon that the sensing MSE deteriorates as the transmit power budget increases under a small sensing performance preference, e.g., $\omega_2=0.2$. {The reason is that partial power is allocated to the side lobes to complete communication tasks, which increases the peak value and beam width of the transmitted waveform at the non-target direction. This increases the sensing MSE between the actual waveform and the ideal waveform.}\par
 In Figs. \ref{fig:antenna communication} and \ref{fig:antenna sensing}, we analyze the effect of performance preference on achieve sum rate and sensing MSE separately under different number of transmit antennas settings. We observe that both C$\&$S performances improve with an increase in the number of transmit antennas for all settings of performance preference. This is because that the multi-antennas providing diversity and spatial multiplexing gains to the ISAC system. As the number of antennas increases, the sensing performance continuously improves in Fig. \ref{fig:antenna communication}, while the communication performance hardly improves further in Fig. \ref{fig:antenna sensing}. For this reason, it can be interpreted as the diversity and spatial multiplexing gains from the multi-antenna have a greater impact on sensing performance.\par
\begin{figure}[t]
	\centering{}\includegraphics[width=3.5in]{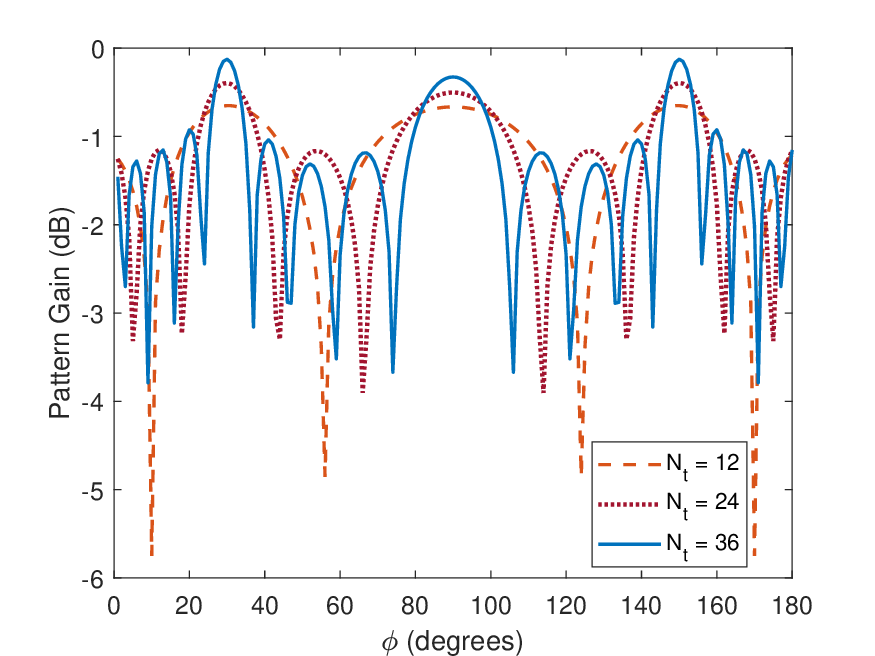}
	\caption{Beam gain of optimized waveforms.}
	\label{fig:beam}
\end{figure}
	{In Fig. \ref{fig:beam}, we analyze the impact of the number of antennas on the transmit waveform, and plot the beam gain of the designed waveforms under different numbers of antennas. With an increase in the number of antennas, it is evident that the width and amplitude of the main lobe towards the target direction become narrower and higher. The reason is that any pattern gain outside target directions is not desired for reducing matching error, i.e., sensing MSE. Additionally, there is a higher level of sidelobe gain in non-target directions such as $40^{\circ} \sim 80^{\circ}$. This is because that high sidelobes are still required to meet the SINR requirements of users outside the sensed target direction.}  

\section{Conclusion}\label{6}

	In this paper, we designed a CI-based ISAC scheme to deal with the MUI in multi-user and multi-target ISAC scenarios. This scheme helped to reduce the power budget for communication, thus enhancing the sensing performance indirectly. To balance the C$\&$S performance, we formulated the waveform design problem as a MOOP. A weighted Tchebycheff-based transformation method was proposed to reformulate this MOOP as a Pareto-optimal problem, resulting in a Pareto front without weakly Pareto optimal solutions. According to the obtained Pareto front, we designed the integrated waveform, under different C$\&$S performance preferences. Finally, we presented simulation results to verify the effectiveness of our proposed scheme.
	
\appendices
	\section{Proof of Lemma \ref{lemma2}}\label{appendixA}
According to objective function \eqref{SOOP1.2a}, we define  
\begin{align}
	g(\widetilde{\mathbf{x}})= \sum^K_{k=1}B\log_2\left(1+\frac{||\mathbf{H}^T_k\widetilde{\mathbf{x}}||^2}{N_0}\right).
\end{align}
To discuss the monotonicity of $g(\widetilde{\mathbf{x}})$, we derive its first-order derivative as
\begin{align}	
	\nabla g(\widetilde{\mathbf{x}})
	= B\log_2\left( \prod_{k=1}^K\frac{N_0+||\mathbf{H}^T_k\widetilde{\mathbf{x}}||^2}{N_0} \right)
	\prod_{k=1}^K\frac{2N_0\mathbf{H}^T_k\widetilde{\mathbf{x}}\mathbf{H}^T_k}{||\mathbf{H}^T_k\widetilde{\mathbf{x}}||^2+N_0}.	
\end{align}
$g(\widetilde{\mathbf{x}})$ is monotonically increasing due to $\nabla g(\widetilde{\mathbf{x}})>0$. Note that holding with equality in inequality \eqref{f1b} when the solution achieve optimal. In this case, we aim to maximize $\mu_k$ instead of directly maximizing the objective function \eqref{33}.

	\section{Proof of Proposition \ref{Prop1}}\label{appendixC}
In Algorithm \ref{algorithm3}, we denote $\left(\mathbf{R}^{(t)},c^{(t)},\mu_k^{(t)},y_l^{(t)}\right)$ as the solution to problem \eqref{17} obtained in the $t$-th iteration, where the objective value is defined as
\begin{align}
	\mathbf{f}'\left(\mathbf{R}^{(t)},c^{(t)},\mu_k^{(t)},y_l^{(t)}\right)=\mathbf{f}\left(\mathbf{R}^{(t)},\mathbf{x}^{(t)}\right).
\end{align}
Since the optimum solution can be attained at each iteration, we have
\begin{align}\nonumber
	\mathbf{f}\left(\mathbf{R}^{(t+1)},\mathbf{x}^{(t+1)}\right)
	=&\mathbf{f}'\left(\mathbf{R}^{(t+1)},c^{(t+1)},\mu_k^{(t+1)},y_l^{(t+1)}\right)\\\nonumber
	\overset{(a)}{\le}&\mathbf{f}'\left(\mathbf{R}^{(t+1)},c^{(t+1)},\mu_k^{(t)},y_l^{(t)}\right)\\
	\overset{(b)}{\le}&\mathbf{f}'\left(\mathbf{R}^{(t)},c^{(t)},\mu_k^{(t)},y_l^{(t)}\right)\\\nonumber
	=&\mathbf{f}\left(\mathbf{R}^{(t)},\mathbf{x}^{(t)}\right),\nonumber
\end{align}
where the inequalities $(a)$ and $(b)$ come from the fact that the sub-objective functions \eqref{44} and \eqref{45} are monotonic, and the constraints \eqref{46b}-\eqref{46d} are linear.

\bibliographystyle{IEEEtran}
\bibliography{ref}

\begin{thebibliography}{10}
\providecommand{\url}[1]{#1}
\csname url@samestyle\endcsname
\providecommand{\newblock}{\relax}
\providecommand{\bibinfo}[2]{#2}
\providecommand{\BIBentrySTDinterwordspacing}{\spaceskip=0pt\relax}
\providecommand{\BIBentryALTinterwordstretchfactor}{4}
\providecommand{\BIBentryALTinterwordspacing}{\spaceskip=\fontdimen2\font plus
\BIBentryALTinterwordstretchfactor\fontdimen3\font minus
  \fontdimen4\font\relax}
\providecommand{\BIBforeignlanguage}[2]{{%
\expandafter\ifx\csname l@#1\endcsname\relax
\typeout{** WARNING: IEEEtran.bst: No hyphenation pattern has been}%
\typeout{** loaded for the language `#1'. Using the pattern for}%
\typeout{** the default language instead.}%
\else
\language=\csname l@#1\endcsname
\fi
#2}}
\providecommand{\BIBdecl}{\relax}
\BIBdecl

\bibitem{conf1}
P.~Wang, Y.~Cao, W.~Ni, and D.~Han, ``Pareto-optimal waveform design for
  multi-user and multi-target {MIMO}-{ISAC} systems,'' in \emph{Proc. IEEE Int.
  Conf. Acoust., Speech, Signal Process. Workshops (ICASSP Workshops)}, Seoul,
  Korea, Apr. 2024.

\bibitem{feng2021joint}
Z.~Feng, Z.~Wei, X.~Chen, H.~Yang, Q.~Zhang, and P.~Zhang, ``Joint
  communication, sensing, and computation enabled 6{G} intelligent machine
  system,'' \emph{IEEE Netw.}, vol.~35, no.~6, pp. 34--42, Nov. 2021.

\bibitem{zhang2021overview}
J.~A. Zhang, F.~Liu, C.~Masouros, R.~W. Heath, Z.~Feng, L.~Zheng, and
  A.~Petropulu, ``An overview of signal processing techniques for joint
  communication and radar sensing,'' \emph{IEEE J. Sel. Topics Signal
  Process.}, vol.~15, no.~6, pp. 1295--1315, Sep. 2021.

\bibitem{chen2021code}
X.~Chen, Z.~Feng, Z.~Wei, P.~Zhang, and X.~Yuan, ``Code-division {OFDM} joint
  communication and sensing system for 6{G} machine-type communication,''
  \emph{IEEE Internet Things J.}, vol.~8, no.~15, pp. 12\,093--12\,105, Feb.
  2021.

\bibitem{mishra2019toward}
K.~V. Mishra, M.~B. Shankar, V.~Koivunen, B.~Ottersten, and S.~A. Vorobyov,
  ``Toward millimeter-wave joint radar communications: A signal processing
  perspective,'' \emph{IEEE Signal Process. Mag.}, vol.~36, no.~5, pp.
  100--114, Sep. 2019.

\bibitem{zhang2021enabling}
J.~A. Zhang, M.~L. Rahman, K.~Wu, X.~Huang, Y.~J. Guo, S.~Chen, and J.~Yuan,
  ``Enabling joint communication and radar sensing in mobile networks—{A}
  survey,'' \emph{IEEE Commun. Surv. Tutorials}, vol.~24, no.~1, pp. 306--345,
  Oct. 2021.

\bibitem{cui2021integrating}
Y.~Cui, F.~Liu, X.~Jing, and J.~Mu, ``Integrating sensing and communications
  for ubiquitous {IoT}: Applications, trends, and challenges,'' \emph{IEEE
  Netw.}, vol.~35, no.~5, pp. 158--167, Nov. 2021.

\bibitem{liu2022survey}
A.~Liu, Z.~Huang, M.~Li, Y.~Wan, W.~Li, T.~X. Han, C.~Liu, R.~Du, D.~K.~P. Tan,
  J.~Lu \emph{et~al.}, ``A survey on fundamental limits of integrated sensing
  and communication,'' \emph{IEEE Commun. Surv. Tutorials}, vol.~24, no.~2, pp.
  994--1034, Feb. 2022.

\bibitem{barneto2019full}
C.~B. Barneto, T.~Riihonen, M.~Turunen, L.~Anttila, M.~Fleischer, K.~Stadius,
  J.~Ryyn{\"a}nen, and M.~Valkama, ``Full-duplex {OFDM} radar with {LTE} and
  5{G} {NR} waveforms: Challenges, solutions, and measurements,'' \emph{IEEE
  Trans. Microw. Theory and Technol.}, vol.~67, no.~10, pp. 4042--4054, Aug.
  2019.

\bibitem{5393298}
C.~R. Berger, B.~Demissie, J.~Heckenbach, P.~Willett, and S.~Zhou, ``Signal
  processing for passive radar using {OFDM} waveforms,'' \emph{IEEE J. Sel.
  Top. Signal Process}, vol.~4, no.~1, pp. 226--238, Jan. 2010.

\bibitem{9399801}
Z.~Cheng, S.~Shi, Z.~He, and B.~Liao, ``Transmit sequence design for
  dual-function radar-communication system with one-bit {DACs},'' \emph{IEEE
  Trans. Wireless Commun.}, vol.~20, no.~9, pp. 5846--5860, Apr. 2021.

\bibitem{huang2020majorcom}
T.~Huang, N.~Shlezinger, X.~Xu, Y.~Liu, and Y.~C. Eldar, ``M{A}{J}o{R}{C}om: A
  dual-function radar communication system using index modulation,'' \emph{IEEE
  Trans. Signal Process.}, vol.~68, pp. 3423--3438, May 2020.

\bibitem{blunt2010embedding}
S.~D. Blunt, M.~R. Cook, and J.~Stiles, ``Embedding information into radar
  emissions via waveform implementation,'' in \emph{Proc. Int. Waveform
  Diversity and Design Conf.}, Niagara Falls, Canada, Sep. 2010, pp. 195--199.

\bibitem{blunt2010intrapulse}
S.~D. Blunt, P.~Yatham, and J.~Stiles, ``Intrapulse radar-embedded
  communications,'' \emph{IEEE Trans. Aerosp. Electron. Syst.}, vol.~46, no.~3,
  pp. 1185--1200, Jul. 2010.

\bibitem{kumari2017ieee}
P.~Kumari, J.~Choi, N.~Gonz{\'a}lez-Prelcic, and R.~W. Heath, ``I{EEE} 802.11
  ad-based radar: An approach to joint vehicular communication-radar system,''
  \emph{IEEE Trans. Veh. Technol.}, vol.~67, no.~4, pp. 3012--3027, Nov. 2017.

\bibitem{barneto2019ofdm}
C.~B. Barneto, L.~Anttila, M.~Fleischer, and M.~Valkama, ``{OFDM} radar with
  {LTE} waveform: Processing and performance,'' in \emph{Proc. IEEE Radio and
  Wireless Symposium (RWS)}, Orlando, USA, May 2019, pp. 1--4.

\bibitem{dong2020low}
F.~Dong, W.~Wang, Z.~Hu, and T.~Hui, ``Low-complexity beamformer design for
  joint radar and communications systems,'' \emph{IEEE Commun. Lett.}, vol.~25,
  no.~1, pp. 259--263, Sep. 2020.

\bibitem{tian2021transmit}
T.~Tian, T.~Zhang, L.~Kong, and Y.~Deng, ``Transmit/receive beamforming for
  {MIMO}-{OFDM} based dual-function radar and communication,'' \emph{IEEE
  Trans. Veh. Technol.}, vol.~70, no.~5, pp. 4693--4708, Apr. 2021.

\bibitem{shi2020joint}
C.~Shi, Y.~Wang, F.~Wang, S.~Salous, and J.~Zhou, ``Joint optimization scheme
  for subcarrier selection and power allocation in multicarrier dual-function
  radar-communication system,'' \emph{IEEE Syst. J.}, vol.~15, no.~1, pp.
  947--958, Apr. 2020.

\bibitem{9124713}
X.~Liu, T.~Huang, N.~Shlezinger, Y.~Liu, J.~Zhou, and Y.~C. Eldar, ``Joint
  transmit beamforming for multiuser {MIMO} communications and {MIMO} radar,''
  \emph{IEEE Trans. Signal Process.}, vol.~68, pp. 3929--3944, Jun. 2020.

\bibitem{10086626}
H.~Hua, J.~Xu, and T.~X. Han, ``Optimal transmit beamforming for integrated
  sensing and communication,'' \emph{IEEE Trans. Veh. Technol.}, vol.~72,
  no.~8, pp. 10\,588--10\,603, Mar. 2023.

\bibitem{10032141}
P.~Gao, L.~Lian, and J.~Yu, ``Cooperative {ISAC} with direct localization and
  rate-splitting multiple access communication: A pareto optimization
  framework,'' \emph{IEEE J. Sel. Areas Commun.}, vol.~41, no.~5, pp.
  1496--1515, Jan. 2023.

\bibitem{9618653}
M.~Temiz, E.~Alsusa, and M.~W. Baidas, ``A dual-function massive {MIMO} uplink
  {OFDM} communication and radar architecture,'' \emph{IEEE Trans. Cognit.
  Commun. Networking}, vol.~8, no.~2, pp. 750--762, Nov. 2022.

\bibitem{10217169}
H.~Hua, T.~X. Han, and J.~Xu, ``{MIMO} integrated sensing and communication:
  {CRB}-rate tradeoff,'' \emph{IEEE Trans. Wireless Commun.}, pp. 1--1, Aug.
  2023.

\bibitem{xiong2023fundamental}
Y.~Xiong, F.~Liu, Y.~Cui, W.~Yuan, T.~X. Han, and G.~Caire, ``On the
  fundamental tradeoff of integrated sensing and communications under gaussian
  channels,'' \emph{IEEE Trans. Inf. Theory}, vol.~69, no.~9, Jun. 2023.

\bibitem{liu2021cramer}
F.~Liu, Y.-F. Liu, A.~Li, C.~Masouros, and Y.~C. Eldar, ``Cram{\'e}r-{R}ao
  bound optimization for joint radar-communication beamforming,'' \emph{IEEE
  Trans. Signal Process.}, vol.~70, pp. 240--253, Dec. 2021.

\bibitem{9468975}
S.~D. Liyanaarachchi, T.~Riihonen, C.~B. Barneto, and M.~Valkama, ``Optimized
  waveforms for 5{G}–6{G} communication with sensing: Theory, simulations and
  experiments,'' \emph{IEEE Trans. Wireless Commun.}, vol.~20, no.~12, pp.
  8301--8315, Jun. 2021.

\bibitem{9965407}
M.~Liu, M.~Yang, H.~Li, K.~Zeng, Z.~Zhang, A.~Nallanathan, G.~Wang, and
  L.~Hanzo, ``Performance analysis and power allocation for cooperative {ISAC}
  networks,'' \emph{IEEE Internet Things J.}, vol.~10, no.~7, pp. 6336--6351,
  Nov. 2023.

\bibitem{9945983}
F.~Dong, F.~Liu, Y.~Cui, W.~Wang, K.~Han, and Z.~Wang, ``Sensing as a service
  in 6{G} perceptive networks: A unified framework for {ISAC} resource
  allocation,'' \emph{IEEE Trans. Wireless Commun.}, vol.~22, no.~5, pp.
  3522--3536, Nov. 2023.

\bibitem{ren2023robust}
Z.~Ren, L.~Qiu, J.~Xu, and D.~W.~K. Ng, ``Robust transmit beamforming for
  secure integrated sensing and communication,'' \emph{IEEE Trans. Commun.},
  vol.~71, no.~9, Jun. 2023.

\bibitem{su2020secure}
N.~Su, F.~Liu, and C.~Masouros, ``Secure radar-communication systems with
  malicious targets: Integrating radar, communications and jamming
  functionalities,'' \emph{IEEE Trans. Wireless Commun.}, vol.~20, no.~1, pp.
  83--95, Sep. 2020.

\bibitem{su2022secure}
N.~Su, F.~Liu, Z.~Wei, Y.-F. Liu, and C.~Masouros, ``Secure dual-functional
  radar-communication transmission: Exploiting interference for resilience
  against target eavesdropping,'' \emph{IEEE Trans. Wireless Commun.}, vol.~21,
  no.~9, pp. 7238--7252, Mar. 2022.

\bibitem{hua2023secure}
M.~Hua, Q.~Wu, W.~Chen, O.~A. Dobre, and A.~L. Swindlehurst, ``Secure
  intelligent reflecting surface aided integrated sensing and communication,''
  \emph{IEEE Trans. Wireless Commun.}, Jun. 2023.

\bibitem{9858656}
K.~Meng, Q.~Wu, S.~Ma, W.~Chen, K.~Wang, and J.~Li, ``Throughput maximization
  for {UAV}-enabled integrated periodic sensing and communication,'' \emph{IEEE
  Trans. Wireless Commun.}, vol.~22, no.~1, pp. 671--687, Jan. 2023.

\bibitem{masouros2015exploiting}
C.~Masouros and G.~Zheng, ``Exploiting known interference as green signal power
  for downlink beamforming optimization,'' \emph{IEEE Trans. Signal Process.},
  vol.~63, no.~14, pp. 3628--3640, May 2015.

\bibitem{6774898}
Y.~Wang, Y.~Liu, Y.~He, X.-Y. Li, and D.~Cheng, ``Disco: Improving packet
  delivery via deliberate synchronized constructive interference,'' \emph{IEEE
  Trans. Parallel Distrib. Syst.}, vol.~26, no.~3, pp. 713--723, Mar. 2015.

\bibitem{jardin2010wideband}
P.~Jardin, F.~Nadal, and S.~Middleton, ``On wideband {MIMO} radar: Extended
  signal model and spectral beampattern design,'' in \emph{Proc. the 7th
  European Radar Conf.}, Paris, France, Nov. 2010, pp. 392--395.

\bibitem{kumari2021adaptive}
P.~Kumari, N.~J. Myers, and R.~W. Heath, ``Adaptive and fast combined
  waveform-beamforming design for {mmWave} automotive joint
  communication-radar,'' \emph{IEEE J. Sel. Topics Signal Process.}, vol.~15,
  no.~4, pp. 996--1012, Apr. 2021.

\bibitem{9022866}
X.~Ma, Y.~Yu, X.~Li, Y.~Qi, and Z.~Zhu, ``A survey of weight vector adjustment
  methods for decomposition-based multiobjective evolutionary algorithms,''
  \emph{IEEE Trans. Evol. Comput.}, vol.~24, no.~4, pp. 634--649, Mar. 2020.

\bibitem{marler2004survey}
R.~T. Marler and J.~S. Arora, ``Survey of multi-objective optimization methods
  for engineering,'' \emph{Struct. Multidiscip. Optim.}, vol.~26, pp. 369--395,
  Mar. 2004.

\bibitem{grant2014cvx}
M.~Grant and S.~Boyd, ``{CVX}: Matlab software for disciplined convex
  programming, version 2.1,'' Mar. 2014.

\bibitem{5447068}
Z.~q. Luo, W.~k. Ma, A.~M.~c. So, Y.~Ye, and S.~Zhang, ``Semidefinite
  relaxation of quadratic optimization problems,'' \emph{IEEE Signal Process
  Mag.}, vol.~27, no.~3, pp. 20--34, Apr. 2010.

\bibitem{10007645}
Y.~Du, Y.~Liu, K.~Han, J.~Jiang, W.~Wang, and L.~Chen, ``Multi-user and
  multi-target dual-function radar-communication waveform design: Multi-fold
  performance tradeoffs,'' \emph{IEEE Trans. Green Commun. Networking}, vol.~7,
  no.~1, pp. 483--496, Mar. 2023.

\end{thebibliography}

\end{document}